\documentclass[twocolumn,trackchanges]{aastex631}
\newcommand{\teff}{$T_{\rm eff}$}
\newcommand{\jz}{$J_{\rm z}$}

\newcommand{\logg}{log$g$}
\usepackage{array}
\usepackage{amsmath} 
\usepackage{tabularx}
\usepackage{chngcntr}
\usepackage{mathtools,amssymb}
\usepackage{comment}

\begin{document}

\title{Ensemble Kinematic Ages for 1.5 Million LAMOST Stars}

\newcommand{\osu}{Department of Astronomy, The Ohio State University, Columbus, 140 W 18th Ave, OH 43210, USA}
\newcommand{\ccapp}{Center for Cosmology and Astroparticle Physics (CCAPP), The Ohio State University, 191 W. Woodruff Ave., Columbus, OH 43210, USA}

\author[0000-0003-4769-3273]{Yuxi(Lucy) Lu}
\affiliation{\osu}
\affiliation{\ccapp}

\author[0000-0002-7549-7766]{Marc H. Pinsonneault}
\affiliation{\osu}
\affiliation{\ccapp}



\begin{abstract}
We present a framework for inferring stellar ages from spectroscopic stellar parameters, calibrated with ensemble kinematics by averaging over the median vertical action, $J_z$, for stars with similar atmospheric and chemical properties, yielding self-consistent age estimates across the Hertzsprung–Russell (HR) diagram.
We refer to these ages as ensemble kinematic ages as the age scale is calibrated from the average kinematics of ensembles of stars with similar stellar parameters. 
Individual stellar kinematics are not used in assigning ages.
We validate the method against subgiant ages, achieving an accuracy of $\sim$30\%, comparable to $[\mathrm{C}/\mathrm{N}]$-based estimates.
We find a clear age–$J_z$ relations that enable age inference up to $\sim$10 Gyr for both the high- and low-$\alpha$ disks.
Applying this framework to 1.5 million LAMOST stars, we derive ages for subgiants and giants with typical uncertainties of $\sim$2 Gyr.
The inferred ages agree well with literature age catalogs, with no significant systematic trends as a function of $\log g$. 
We also demonstrate the potential of empirical isochrones to calibrate theoretical stellar models.
We identify an old ($\sim$7 Gyr) population within the low-$\alpha$ disk but draw no firm conclusions.
Although ensemble kinematic ages are statistical and sensitive to selection effects, Galactic potential assumptions, and Galactic location, they provide a robust population-level tool for Galactic archaeology, complementing traditional age indicators and extending age estimates across the full HR diagram.
\end{abstract}



\section{Introduction} \label{sec:intro}
Age-dating stars has become a rapidly growing field over the past decade, driven by large, full-sky photometric and spectroscopic surveys such as Gaia \citep{gaia}, Kepler \citep{kepler}, K2 \citep{K2}, TESS \citep{TESS}, ZTF \citep{ztfdata,ztftime}, APOGEE \citep{apogee}, and GALAH \citep{galah}. 

A key underlying assumption common to many modern stellar age methods is that stars occupying a small region of stellar-parameter space (e.g., $T_{\mathrm{eff}}$, $\log g$, and chemical abundances) are drawn from a relatively compact distribution of ages, reflecting the fact that stellar parameters evolve systematically with time. 
This assumption is supported empirically by multiple independent dating techniques showing that stellar parameters encode significant age information, even if with intrinsic scatter.
For example, high-precision distances and photometry from Gaia, together with accurate chemical abundances (particularly metallicity) from large spectroscopic surveys such as LAMOST \citep{lamost}, have enabled isochrone fitting of subgiants with median age uncertainties of less than 10\% \citep[e.g.,][]{Xiang2022}, reflecting the rapid evolution of stellar parameters in this phase of stellar evolution.
Large-scale time-domain surveys such as Kepler, TESS, and ZTF have enabled measurements of large numbers of stellar rotation periods for dwarfs and oscillation frequencies for giants \citep[e.g.,][]{Pinsonneault2018,pinsonneault2024,Santos2021,McQuillan2014,Claytor2020,Holcomb2022,Colman2024,Lu2022_prot,Silva2017}. 
When combined with photometric and abundance information, these observables can be used to infer stellar ages via gyrochronology \citep[e.g.,][]{Barnes2010} and asteroseismology \citep[e.g.,][]{Ulrich1986_scaling_relation1,Brown1991_scaling_relation2,Kjeldsen1995_scaling_relation3}, further demonstrating that stellar parameters tightly correlate with stellar age.
In addition, stars mix material processed during the CNO cycle to their surfaces as they ascend the lower giant branch during the first dredge-up phase \citep{Iben1967}, producing predictable, mass-dependent changes in surface abundances. 
As a result, the carbon-to-nitrogen ratio ([C/N]) can also be used as an age indicator for giant stars \citep[e.g.,][]{Martig2016, Ness2016, stonemartinez2025, Roberts2025}, again reinforcing the idea that chemical abundance patterns contain strong statistical information about stellar age.

With current and near-future surveys such as Milky Way Mapper \citep{MWM}, LSST \citep{LSST}, and Roman \citep{Spergel2015}, stellar ages are expected to remain among the most important stellar parameters for advancing our understanding of stellar physics, planet evolution, and galaxy formation.

However, one age indicator that has received comparatively little attention, despite recent advances in astrometric surveys such as Gaia, is stellar kinematics. Although 6-D phase-space information has long been used to estimate the ages of young associations and clusters by tracing their members back to a common origin \citep[e.g.,][]{Blaauw1964,Ferraro2012,Kuhn2019}, kinematic ages based on the age--velocity relation \citep[AVR; e.g.,][]{Spitzer1951,Lacey1984,sellwood2014,Bird2013} have largely been limited to coarse classifications of stellar populations as young or old. 
The velocity dispersion, particularly in the vertical direction ($\sigma_{v_z}$), is known to increase with age, likely due to both stochastic orbital scattering by giant molecular clouds \citep[e.g.,][]{Spitzer1951} and upside-down formation scenarios \citep[e.g.,][]{Bird2013}, in which galactic disks form kinematically hot and subsequently cool over time. 
Another kinematic tracer of disk heating is the vertical action, $J_z$, defined as the phase-space area enclosed by a star's vertical oscillation.
An important advantage of using $J_z$ over $\sigma_{v_z}$ is that it can be measured on a star-by-star basis and is an adiabatic invariant under gradual changes in the Galactic potential, such as the growth of the Milky Way (MW).

Despite extensive studies of the MW's heating history because of its implications for galaxy formation \citep[e.g.,][]{Ting2019}, kinematics have rarely been used as a primary stellar age indicator \citep[e.g.,][]{Newton2016,Hamer2020,Schmidt2026}, as kinematic ages are often considered to be imprecise. 
This concern is well justified for individual stars, for which kinematic age estimates typically have uncertainties of several gigayears \citep[e.g.,][]{Sagear2024}. 
However, when stars can be grouped into populations that are plausibly coeval, their median or mean kinematic properties can provide a robust estimate of the population's average age. 
For example, \citet{Angus2020} and later \citet{Lu2021_kinage} inferred kinematic ages for field dwarfs by grouping stars with similar effective temperatures ($T_{\rm eff}$) and rotation periods, under the assumption that stars sharing similar temperatures and rotation periods are of similar age.
These kinematic ages successfully reproduce the period--temperature relations of open clusters and have been used to calibrate gyrochronology beyond 4~Gyr, as well as to study stellar spin-down at old ages \citep[e.g.,][]{Lu2026,Lu2024,See2024,Lu2024_Mdwarf}. 
This result is not unexpected: age estimates based on stellar parameters implicitly assume that stars sharing a particular set of observables have a characteristic or typical age. 
In this sense, inferring an average kinematic age for stars with similar properties is conceptually analogous to age-dating stars based on their shared stellar observables.

One important caveat in using kinematic information as an age indicator is that it assumes smooth, secular dynamical evolution. 
It does not account for non-secular processes such as disk heating from satellite mergers or abrupt shifts in star formation conditions, which can alter stellar kinematics independently of age.
Although the MW disk likely became rotationally supported at early times, prior to the formation of the high-$\alpha$ disk ($\sim$12–13 Gyr ago, at [Fe/H] $\sim -$1; \citealt{Xiang2022, Belokurov2022, Conroy2022}), the clear chemical distinction between the high- and low-$\alpha$ populations may suggest they formed during different evolutionary phases. 
In particular, the high-$\alpha$ disk likely formed in a dynamically hot, turbulent environment \citep[e.g.,][]{pinsonneault2024, Xiang2022}, where stellar kinematics were imprinted at birth and subsequently evolved less through secular heating.
As a result, the high-$\alpha$ population may exhibit a weaker kinematic–age relation than the more quiescently formed low-$\alpha$ disk.
Moreover, perturbations such as close-encounters with satellite galaxies can also heat up the disk independent of age \citep[e.g.,][]{Velazquez1999}. 
However, despite the causes, as long as an kinematic-age relation exist, kinematic can be used as an age-indicator.

In this paper, we test the accuracy and limitations of kinematic age estimates by comparing them with subgiant ages. 
In particular, we investigate an approach similar to that of \citet{Lu2021_kinage}, in which, for each star, the age is assigned as the average kinematic age of stars with similar stellar parameters, such as $T_{\rm eff}$ and absolute magnitude.
We refer to these as ensemble kinematic ages. 
In \autoref{sec:datamethod}, we describe the data and methodology used in this study.
In \autoref{sec:kinage}, we assess the accuracy and limitations of the inferred kinematic ages through understanding the kinematic-age relation in the high- and low-$\alpha$ disk separately and through comparisons with subgiant ages. 
Finally, in \autoref{sec:results}, we infer ensemble kinematic ages for 1.5 million LAMOST stars based on $T_{\rm eff}$, $\log g$, and 10 selected elemental abundances (C, Mg, Al, Si, Ca, Ti, Cr, Mn, Ni, and Ba), and demonstrate several potential applications of this age catalog. 
We discuss limitations and caveats associated with the use of these ages in \autoref{sec:limitation}.

\section{Data \& Methods} \label{sec:datamethod}
\subsection{Data}
We use the subgiant age catalog of \citet{Xiang2022}, which provides isochrone-based ages and absolute magnitudes in the 2MASS $K$ band ($M_K$), as our comparison sample. 
The subgiant ages have a median uncertainty of 7\%. 
Isochrone ages for subgiants are particularly precise because their effective temperatures ($T_{\rm eff}$) and luminosities evolve rapidly across the HR diagram. 
This sample is used to calibrate the age--$\ln J_z$ relation and to evaluate the accuracy and limitations of our ensemble-kinematically calibrated stellar ages.

Additionally, we use detailed abundances, $T_{\rm eff}$, and $\log g$ from the LAMOST DR5 value-added catalog \citep{lamost,Xiang2019}, derived with the data-driven Payne model \citep{Ting2017,Ting2019b}. 
After validating our methodology, we apply the calibrated relation to this larger sample to infer stellar parameter ages calibrated with ensemble kinematics.

The 6-D kinematic properties, including Galactocentric velocities ($v_R$, $v_\phi$, $v_z$), are calculated from Gaia DR3 measurements (RA, Dec., parallax, proper motions, and radial velocity) \citep{Gaiadr3} by transforming from the solar-system barycentric ICRS reference frame to Galactocentric Cartesian and cylindrical coordinates using \texttt{astropy} \citep{astropy:2013,astropy:2018,astropy2022}, with updated solar motion parameters from \citet{Hunt:2022}. 
The actions ${\bf J} = (J_R, J_\phi, J_z)$ and guiding radius ($R_g$) are calculated using the \texttt{MilkyWayPotential2022} in \texttt{gala} \citep{gala2017}, with the additional constraint that the circular velocity at the solar position is 229~km~s$^{-1}$ \citep{Eilers2019}. 
Actions are computed using the ``St\"ackel Fudge'' \citep{Binney2012,Sanders2012} as implemented in \texttt{galpy} \citep{galpy}.
The subgiant sample is used to calibrate the age--$\ln J_z$ relation, which translates ensemble kinematic measurements into stellar ages.

\begin{figure*}[ht!]
\includegraphics[width=\textwidth]{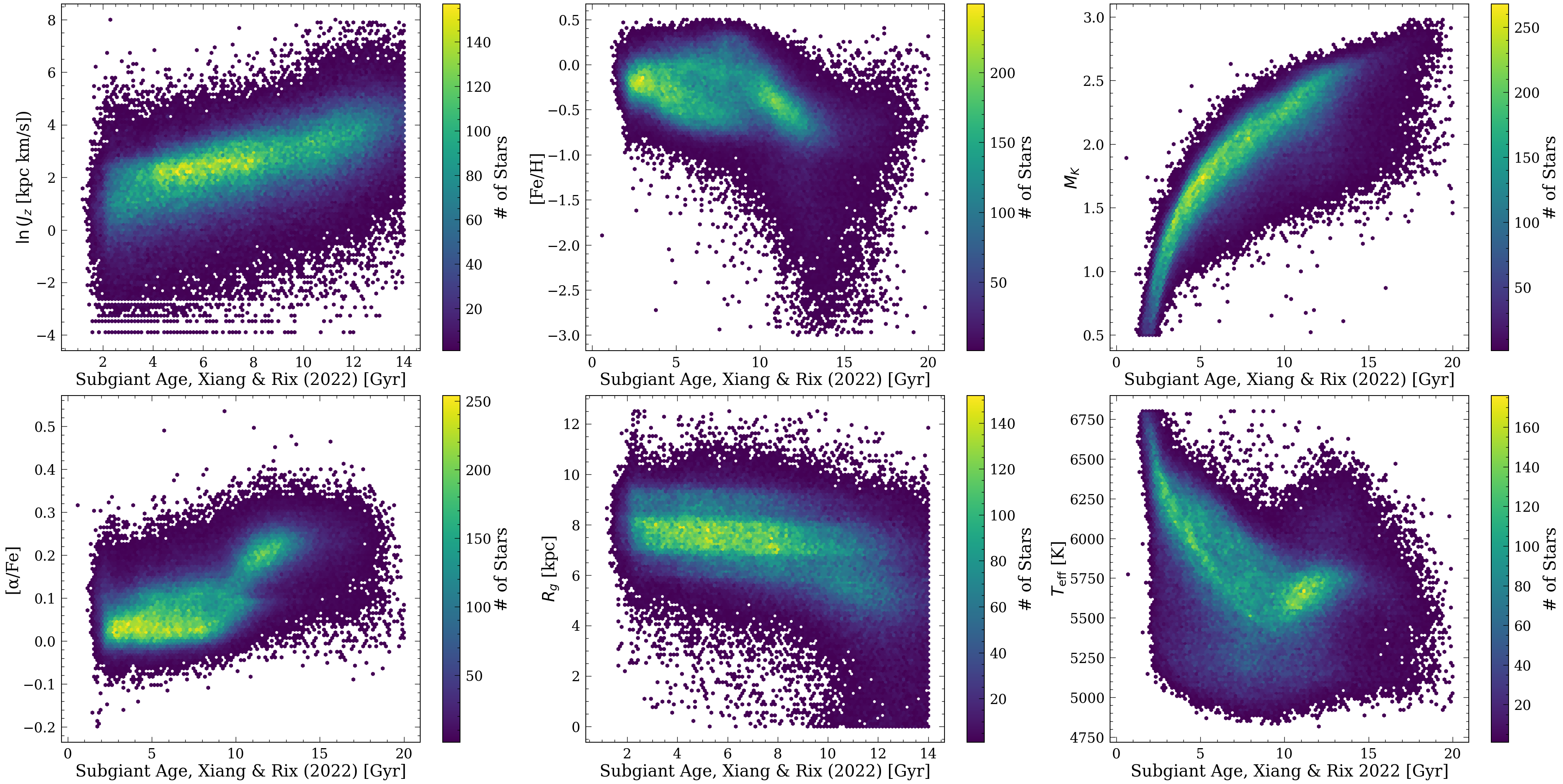}
\caption{Relation between the subgiant ages \citep{Xiang2022} and vertical action ($J_z$), LAMOST DR5 [Fe/H], [$\alpha$/Fe], $T_{\rm eff}$ \citep{Xiang2019}, absolute magnitude in 2MASS $K$ band ($M_K$) \citep{Xiang2022}, and guiding radius ($R_g$). The stellar parameters on the $y$-axis are used to infer ensemble kinematic ages. A clear relation between age and $J_z$ is evident.}
\label{fig:fig1}
\end{figure*}

\autoref{fig:fig1} shows the relation between subgiant ages \citep{Xiang2022} and various stellar parameters used to infer ensemble kinematic ages. 
A clear relation between age and $J_z$ is apparent.

\subsection{Method: Ensemble Kinematic Ages}\label{subsec:method}
As described in the introduction, inferring ensemble kinematic ages involves first grouping stars that are similar in stellar parameters, measuring their average kinematic properties, and then applying an age--kinematic relation to infer age. 
Since $J_z$ is less sensitive to gradual changes in the Milky Way potential compared to vertical velocities, we use $\ln J_z$ as our kinematic age indicator.

To test this method, we first adopt the parameters reported in \citet{Xiang2022}. 
For a fair comparison, we use the same stellar parameters ($T_{\rm eff}$, $M_K$, [Fe/H], and [$\alpha$/Fe]) employed in their subgiant age determination.
For a target star, we measure the median $\ln J_z$ of stars whose stellar parameters are consistent with those of the target within the typical measurement uncertainties. 
We adopt the measurement uncertainties to define a local ensemble of stars that are observationally indistinguishable in stellar-parameter space, rather than to optimize predictive accuracy.
The uncertainties are 30~K, 0.08~mag, 0.07~dex, and 0.03~dex for $T_{\rm eff}$, $M_K$, [Fe/H], and [$\alpha$/Fe], respectively.
Each selected group is required to contain between 20 and 50 stars: the lower limit ensures statistical significance, while the upper limit maintains smaller bin sizes in high-density regions where stellar parameters vary slowly. 
We optimized the bin sizes using a grid search over minimum bin sizes of 10, 20, 50, 100, and 200 stars, and maximum bin sizes of 50, 100, 300, 400, and 500 stars, while keeping the abundance bin sizes fixed. 
The optimal bin sizes were selected by minimizing the reduced $\chi^2$ relative to a randomly selected sample of 100 subgiant ages from \cite{Xiang2022}. 
We find that the inferred ages are largely insensitive to the choice of bin sizes down to a minimum bin size of 10, indicating that our results are robust to this parameter.
If a group does not satisfy this criterion, the bin size for each parameter is increased or decreased by 10\% iteratively. 
If adaptive bin sizes exceed 200\% of the original selection, the bin is fixed at 200\% when defining the group.

Finally, the median $\ln J_z$ of the selected stars is converted to an ensemble kinematic age using a calibrated $J_z$--age relation. 
This age is assigned to the target star. The process is repeated until all stars in the sample have an ensemble kinematic age.

This method assumes:
\begin{enumerate}
\item Stars with similar atmospheric and chemical parameters are approximately coeval.
\item The intrinsic age spread within each stellar-parameter ensemble is sufficiently small that the ensemble age is representative of its members.
\item The age--$\ln J_z$ relation is monotonic and can be parameterized.
\item The calibrated age--$\ln J_z$ relation is broadly applicable across the stellar populations analyzed.
\item Contributions to vertical heating other than age and metallicity average out within stellar-parameter ensembles.
\end{enumerate}

\subsection{The $\ln J_z$-age relation for the high- and low-$\alpha$ disks}
As noted in the introduction, the distinct chemical signatures of the high- and low-$\alpha$ disks may reflect differences in their formation and evolutionary histories, which could translate into systematically different kinematic behavior. 
We are now in a position to test differences in the kinematic properties of these two populations using precise subgiant ages.

\begin{figure*}
\includegraphics[width=\textwidth]{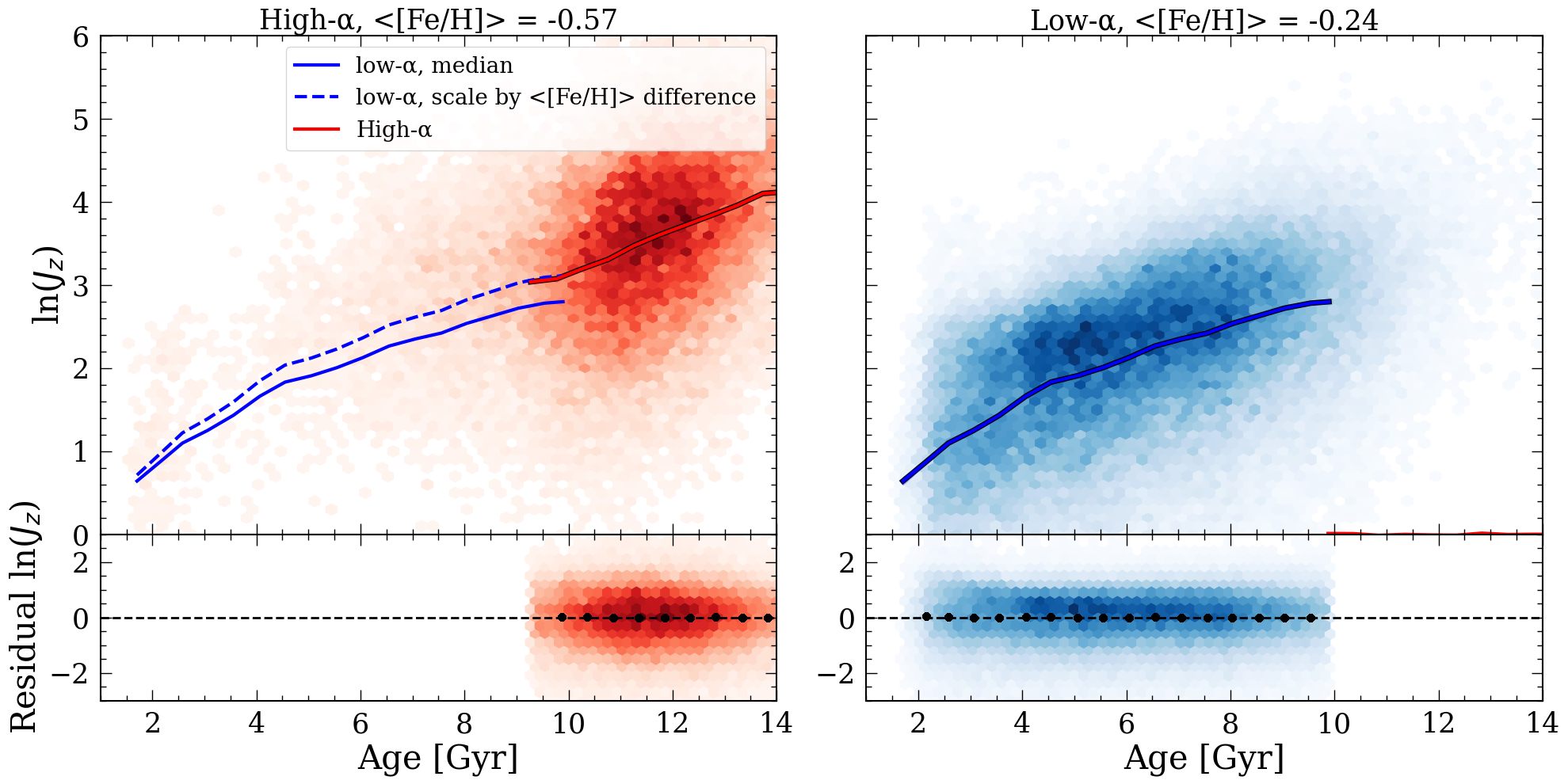}
\caption{The $\ln J_z$–age relation for the high-$\alpha$ (left) and low-$\alpha$ (right) Galactic disk populations, using stellar ages from \citet{Xiang2022}. 
The bottom histograms show the residual after subtracting the median trends and the black points show the running median of the residual.
The red and blue solid curves denote the running medians for the high- and low-$\alpha$ disks, respectively, with shaded regions indicating the median absolute deviation (MAD) about each median trend. 
The offset between the two $\ln J_z$–age relations is consistent with the metallicity difference between the populations, as shown by the dashed blue line in the left plot, where we scaled the low-$\alpha$ relation by the metallicity dependence shown in \autoref{eq1}.
However, the slopes for the two disks are slightly different, indicating different formation scenarios.}
\label{fig:fig2_1}
\end{figure*}

\autoref{fig:fig2_1} presents the $\ln J_z$–age relations for the two populations separately, where the high- and low-$\alpha$ disks are defined as in \cite{Xiang2022}.
We only fit to the bulk of the age population for each disk, which for the high-$\alpha$ disk is $>$ 9 Gyr, and for the low -$\alpha$ disk is $<$ 10 Gyr. 
The blue and red curves show the running medians for the high- and low-$\alpha$ disks, respectively. 
The shaded regions indicate $1.5\times$ the median absolute deviation (MAD), normalized by the square root of the number of stars about each trend, and are thinner than the corresponding lines.

\begin{figure*}[ht!]
\includegraphics[width=\columnwidth]{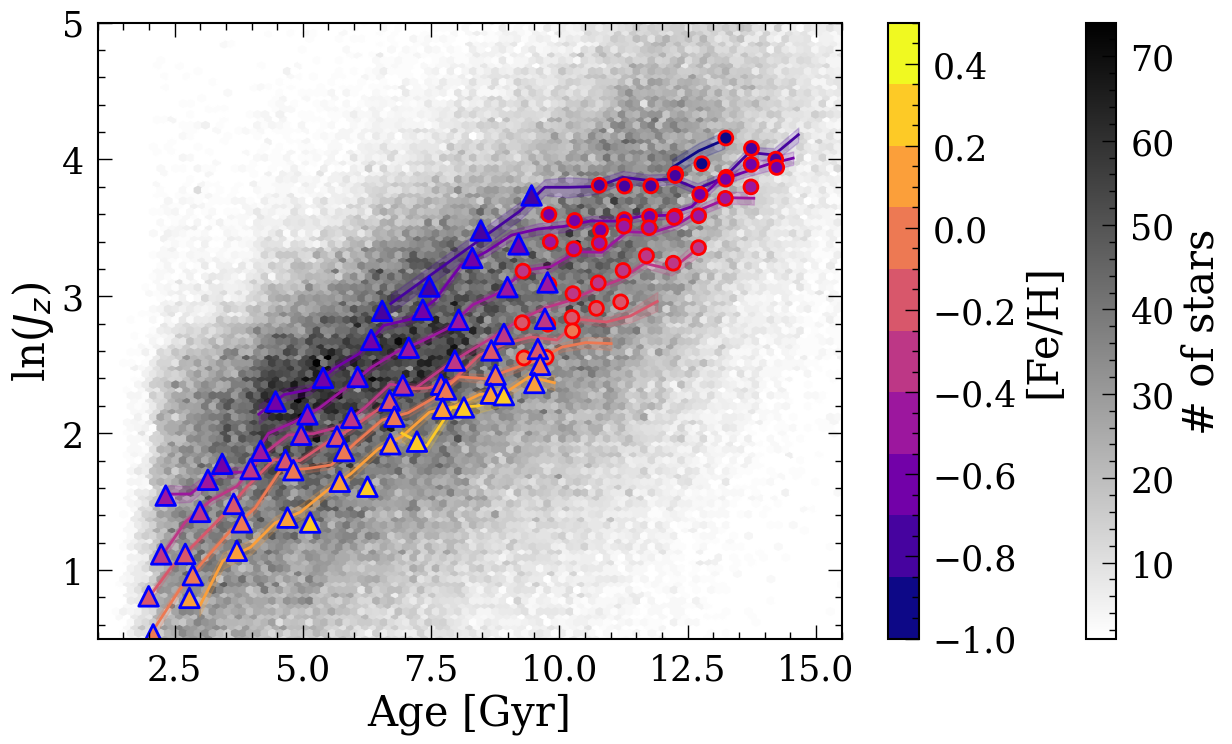}
\includegraphics[width=\columnwidth]{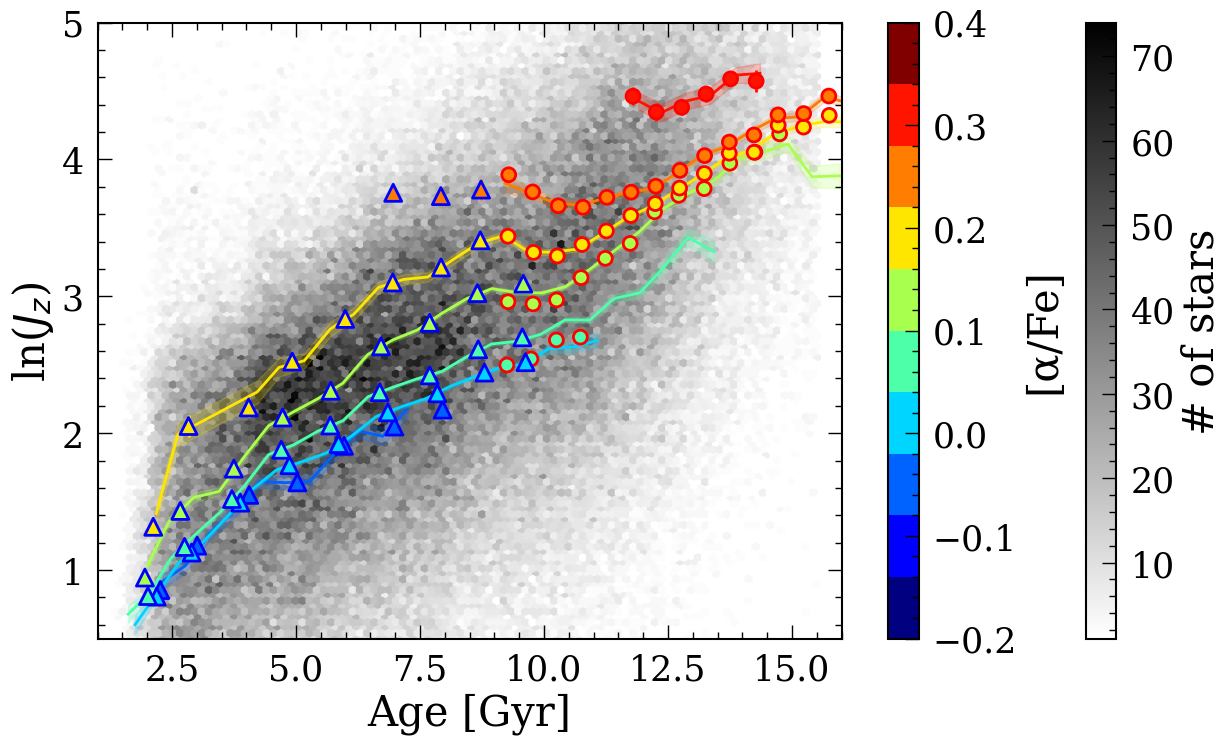}
\caption{The background gray histogram shows the $\ln J_z$–age relation (same as the top-left panel of \autoref{fig:fig1}). 
Red-outlined points and blue-outlined triangles show the moving median of $\ln J_z$ as a function of age in bins of metallicity (left) or [$\alpha$/Fe] (right), indicated by color, for the high- and low-$\alpha$ disks, respectively. 
The colored lines show the corresponding $\ln J_z$–age relations at different metallicities when all stars are considered together, without separating the two populations.
Although a different relation may exist at ages $\gtrsim 10,\mathrm{Gyr}$, coinciding with the onset of the high-$\alpha$ disk, the solid lines closely trace the data points, indicating a continuous $\ln J_z$–age relation when separating by [Fe/H] or [$\alpha$/Fe], independent of disk membership.}
\label{fig:fig2_2}
\end{figure*}

As expected, an offset and a change in slope exists between the median $\ln J_z$–age relations of the two disks, indicating different heating histories. 
However, the two populations also differ in their mean metallicities, where the high-$\alpha$ population has a mean [Fe/H] of $-0.56$ dex, while the low-$\alpha$ population has a mean of $-0.24$ dex. 
The negative metallicity gradient in the Milky Way implies that higher-metallicity stars are more likely to reside in the inner Galaxy, where the vertical restoring force is stronger, leading to less heating and smaller $J_z$ values \citep{Ting2019}. 
Moreover, differences in metallicity correspond to differences in the birth radii of these stars \citep[e.g.,][]{Minchev2018, Lu2024_rb}, which may also imply differences in their birth vertical actions.
The metallicity dependence can be seen also in the left plot of \autoref{fig:fig2_2}.
Given the metallicity dependence of the $\ln J_z$–age relation, this difference predicts a $\sim$10\% higher $\ln J_z$ at fixed age, consistent with the observed offset (indicated by the dashed blue line in the left panel of \autoref{fig:fig2_1}).
After subtracting the median trends, the dispersions around the median are 1.17 and 1.14 dex for the high- and low-$\alpha$ disks, respectively.
The similar residual scatter further supports the idea that a continuous $\ln J_z$–age relation can be applied to both disks. 
We also examined the relation between vertical velocity dispersion and age and found similar results.

We further investigate whether a universal $\ln J_z$–age relation, decomposed in [Fe/H] or [$\alpha$/Fe], can be applied across both disks. 
To this end, we measure the $\ln J_z$–age–[Fe/H] relation separately for the high- and low-$\alpha$ disks and also compute the relation for the full sample without distinguishing between the two populations. 
\autoref{fig:fig2_2} shows the results: red-outlined points and blue-outlined triangles represent the running medians for the high- and low-$\alpha$ disks, respectively, with colors indicating the [Fe/H] (left) or [$\alpha$/Fe] (right) bins. 
The solid colored lines show the corresponding relations for the combined sample.
We find a continuous relation across the full sample. 
Although a possible transition exists between the two disks, with older stars exhibiting a shallower slope that depends on metallicity, the transition is smooth and can be described by a global $\ln J_z$–age–[Fe/H] relation without explicitly separating stars into high- and low-$\alpha$ populations.
Consequently, we fit a global $\ln J_z$–age–[Fe/H] relation, as described in the next section, to translate ensemble kinematics into stellar ages.

Previous studies have shown that the two disks exhibit smooth and continuous changes in structural and kinematic properties, such as scale length and vertical velocity dispersion \citep{Bovy2013}, and display similar levels of intrinsic elemental abundance dispersion \citep[e.g.,][]{Lu2022_disp}. 
The higher velocity dispersion often reported for the high-$\alpha$ disk, or for stars older than $\sim$8 Gyr in the solar neighborhood \citep[e.g.,][]{Mackereth2019, Sharma2021}, may partly arise from selection effects and age uncertainty. 
This smooth transition can also be seen in \autoref{fig:fig2} right plot, where we plotted the median $\ln J_z$–age relations at different [$\alpha$/Fe] instead of [Fe/H] as shown in the left. 

Future work should investigate the chemo-dynamic history of the MW further but in this paper, we find a common $\ln J_z$–age-[Fe/H] relation exist among the high- and low-$\alpha$ disk.

\subsection{Fitting the ln\jz-age relation}
Since different age catalogs and methods can yield ages on different systematic scales \citep[e.g.,][]{Lu2026}, we refit the $\ln J_z$--age relation to match the subgiant ages. 
For comparison, using the guiding-radius-dependent mean $J_z$--age relation from \citet{Ting2019} can reproduce subgiant ages with similar variance, albeit with a larger bias than the refitted relation described below. 
This bias is unsurprising, as \citet{Ting2019} used spectroscopic ages calibrated with asteroseismic ages from APOKASC--2 \citep{Pinsonneault2018}, which likely have an offset relative to the subgiant ages (see \autoref{fig:fig6}).

To refit the $\ln J_z$--age relation, we first computed the moving median of $\ln J_z$ in age, using a window of 0.3~Gyr and a sliding step of 0.1~Gyr. 
We then fit a 5$^{\rm th}$-order polynomial to the running median. 
Because the age--velocity relation is known to depend on metallicity or location of the Galaxy \citep[e.g.,][]{Sharma2021}, we also measured the moving median of $\ln J_z$ in age for metallicities between $-0.8$~dex and $0.2$~dex in steps of $\sim0.11$~dex. 
The metallicity dependence was incorporated by scaling the 5$^{\rm th}$-order polynomial linearly in [Fe/H] to minimize $\chi^2$ relative to the moving medians. 
The resulting $\ln J_z$--age relation is:

\begin{multline}\label{eq1}
    \tau (\overline{\ln{J_z}}, {\rm [Fe/H]}) = \\
    (-9.9 + 30.8\,\overline{\ln{J_z}} - 30.6\,\overline{\ln{J_z}}^2 + 15.1\,\overline{\ln{J_z}}^3 - \\
    3.4\,\overline{\ln{J_z}}^4 + 0.3\,\overline{\ln{J_z}}^5)(0.34\,{\rm [Fe/H]} + 1.16),
\end{multline}

where $\tau$ is the stellar age and $\overline{\ln J_z}$ is the median $\ln J_z$ measured using the method described in the previous section.

\begin{figure}
\includegraphics[width=\columnwidth]{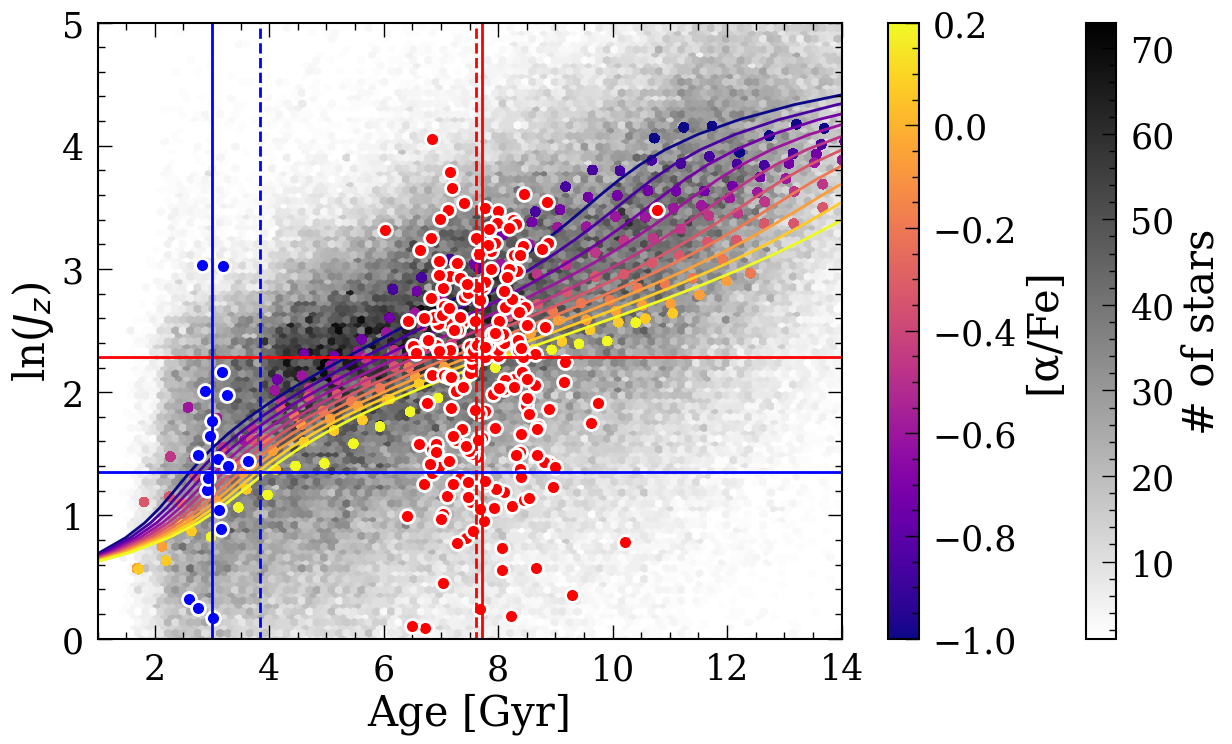}
\caption{Similar to \autoref{fig:fig2} but over-plotting the best-fit model.
Colored points show the moving median of $\ln J_z$ in age at various metallicities, and the colored lines show the refitted $\ln J_z$--age relation at the same metallicities following \autoref{eq1}. 
Red and blue points and lines illustrate two examples of obtaining ensemble kinematic ages: blue corresponds to ($T_{\rm eff}$, $M_K$, [Fe/H], [$\alpha$/Fe]) = (5500 K, 1~mag, -0.1~dex, 0.1~dex), and red to (5169~K, 1.8~mag, 0~dex, 0~dex).
White-outlined points indicate selected stars with parameters consistent with the target star. 
Horizontal solid lines show the median $\ln J_z$ for each group. 
Vertical solid lines show the true subgiant ages, and vertical dashed lines show the ensemble kinematic ages for the target stars. 
The red vertical dashed line is hidden under the solid line. 
Ensemble kinematic ages agree within $\sim1$~Gyr of the median subgiant ages for stars with matching parameters.}
\label{fig:fig2}
\end{figure}

\autoref{fig:fig2} illustrates the $\ln J_z$--age relations at various metallicities, along with two examples of deriving ensemble kinematic ages. 
Blue points and lines correspond to a star with $(T_{\rm eff}, M_K, {\rm [Fe/H]}, [\alpha/{\rm Fe}]) = (5500~{\rm K}, 1~{\rm mag}, -0.1~{\rm dex}, 0.1~{\rm dex})$, and red points and lines correspond to $(5169~{\rm K}, 1.8~{\rm mag}, 0~{\rm dex}, 0~{\rm dex})$. 
White-outlined points indicate stars whose parameters agree with the target star within uncertainties. 
Horizontal solid lines show the median $\ln J_z$ for each group, vertical solid lines indicate the median subgiant ages for each group, and vertical dashed lines mark the ensemble kinematic age of the targeted stars inferred from the $\ln J_z$--age relation.

\section{Ensemble Kinematic Ages for Subgiants}\label{sec:kinage}
\begin{figure}
\includegraphics[width=\columnwidth]{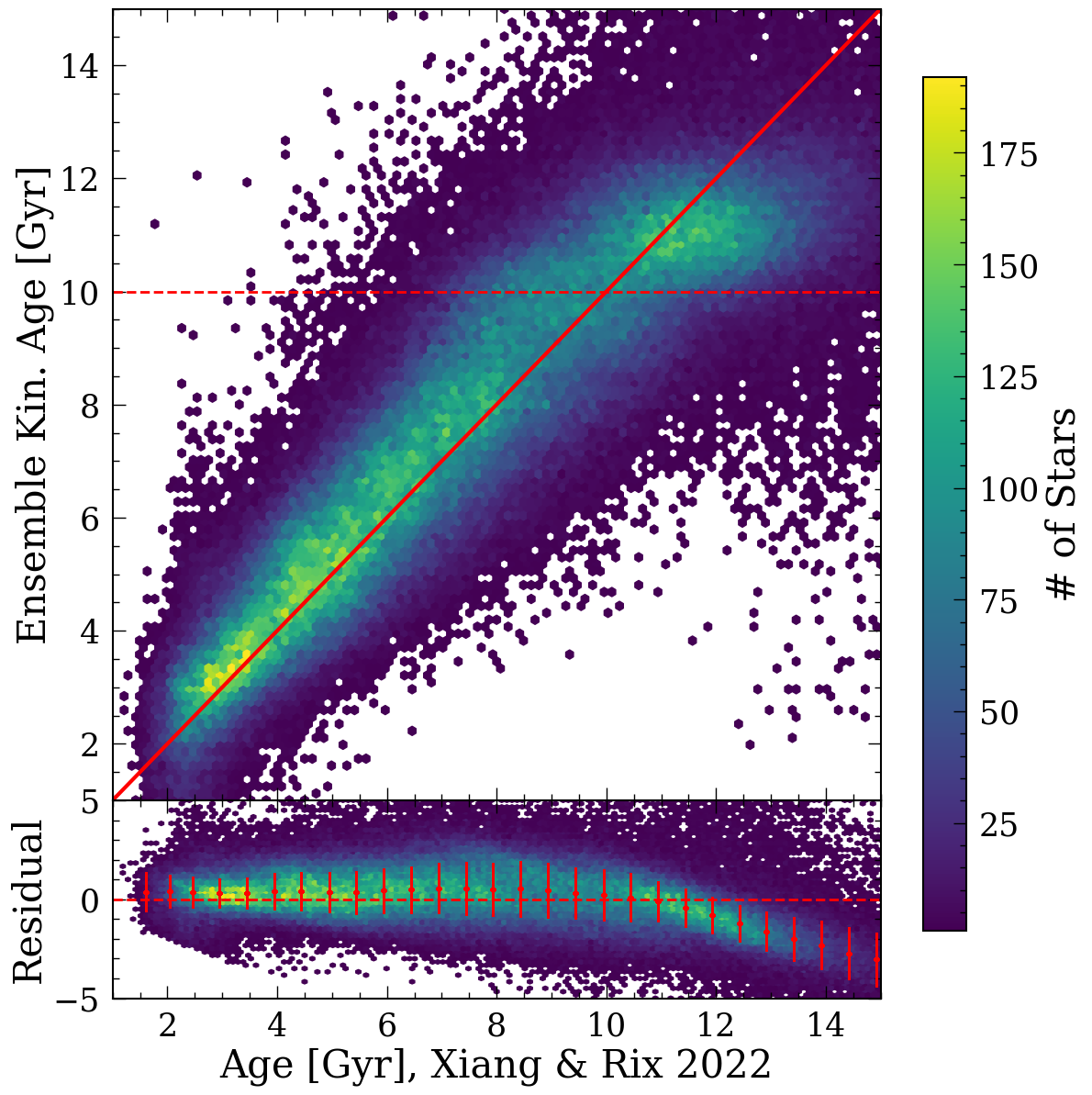}
\caption{Comparison of ensemble kinematic ages with subgiant ages. 
The bottom panels show the residuals, with the running median indicated by red points.
The error bars on the red points represent the 1.5*MAD. 
Ensemble kinematic ages reproduce subgiant ages with a median difference of 0.9~Gyr and a MAD of 2.6~Gyr ($\sim$30\%), comparable to the precision of [C/N]-based ages. 
An increase in bias is apparent for ensemble kinematic ages above 10~Gyr (dashed horizontal line in the top panel), likely due to our $\ln J_z$ relation not fully capturing stars older than 10~Gyr, particularly at [Fe/H] $< -0.5$ (see \autoref{fig:fig2}).}
\label{fig:fig3}
\end{figure}

Following the method described in the previous section, we inferred ensemble kinematic ages for $\sim$200,000 stars from \citet{Xiang2022}, after excluding stars with ensemble kinematic ages $>$ 15~Gyr or $<$ 0~Gyr. 
Subgiant ages provide not only a sample of accurate stellar ages for studying kinematic age relations, but also an independent validation set to test our method.
\autoref{fig:fig3} (top panel) shows the comparison for the full sample. 
The ensemble kinematic ages reproduce the subgiant ages from \citet{Xiang2022} with a median difference (MED), calculated by Median(ensemble kinematic ages $-$ subgiant ages), of 0.9~Gyr and a median absolute deviation (MAD), calculated by Median($|$ensemble kinematic ages-Median(subgiant\ ages)$|$), of 2.6~Gyr (or $\sim$30\%), comparable to [C/N]-based ages \citep{stonemartinez2025,Roberts2025,Lu2026}. 

A clear increase in bias is observed for subgiant ages above 12.5~Gyr, or for ensemble kinematic ages exceeding 10~Gyr (dashed horizontal line in the top panel; see the next section for details).
The bias likely arises because our $\ln J_z$ function does not adequately capture the behavior of high-$\alpha$ stars older than 10~Gyr (see \autoref{fig:fig2}).
The bias also reflects the flattening of the running median in the $\ln J_z$–age relation shown in \autoref{fig:fig2}, particularly for [Fe/H] $< -0.5$. 
Future work could refine the $\ln J_z$–age relation to provide tighter constraints on stellar ages.

Our tests on subgiant age recovery indicate that ensemble kinematic ages achieve an accuracy of approximately 30\%, comparable to [C/N]-based estimates, for typical disk stars. 
With the forthcoming Gaia DR4, which will roughly double the number of stars with full 6-D phase-space information, ensemble kinematic ages are poised to become a powerful tool for Galactic archaeology and age calibration.

\subsection{Parameter-Space Regions Where Ensemble Kinematic Ages Break Down}\label{subsec:suitablespace}
\begin{figure*}
\includegraphics[width=\textwidth]{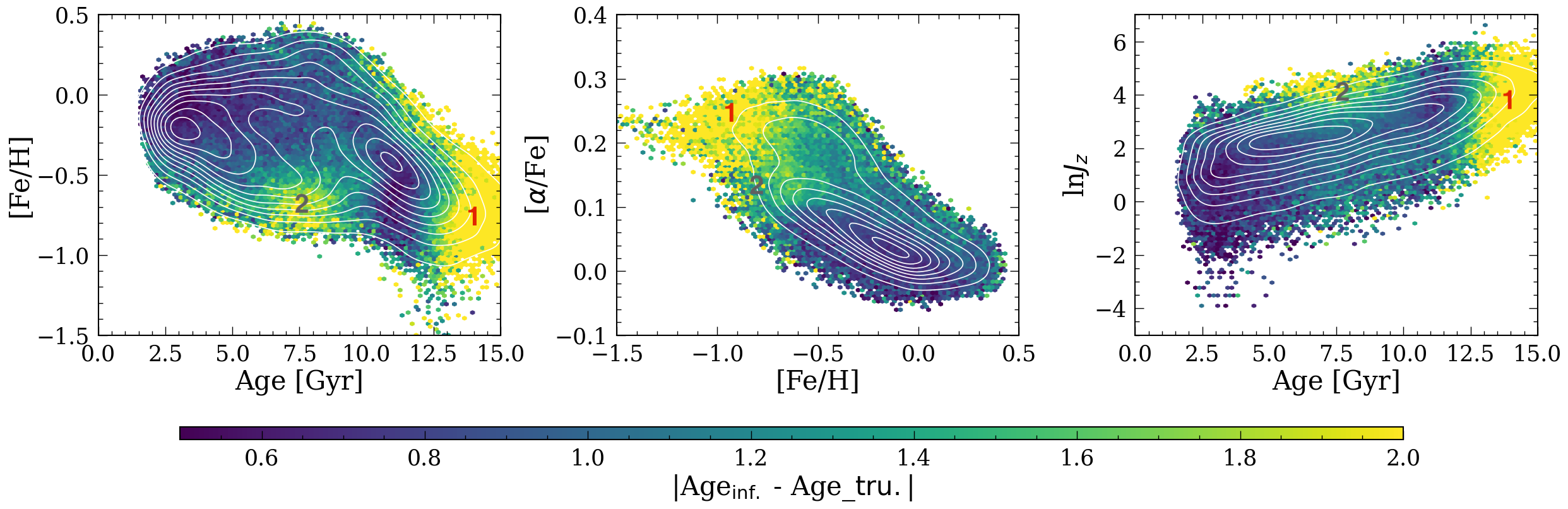}
\caption{Age--metallicity relation (left), the [Fe/H]--[$\alpha$/Fe] plane (middle), and the $\ln J_z$--age relation (right), colored by the average absolute difference between subgiant ages and ensemble kinematic ages. The age on the $x$-axis in the left and right panels corresponds to the subgiant age. 
White contours show kernel density estimation (KDE) densities computed using \texttt{SciPy}.
The regions where the $\ln J_z$--age relation does not perform well include ages $>$ 12.5~Gyr (red ``1''), corresponding to epochs before the formation of the rotationally supported disk, and ages between $\sim6$ and 8~Gyr at [Fe/H] $\sim -0.75$~dex (dark gray ``2''), corresponding to the metal-poor end of the low-$\alpha$ disk that could be related to the GSE merger. The latter population shows elevated $J_z$, suggesting a more dynamically turbulent formation or heating history.}
\label{fig:fig4}
\end{figure*}

During the process of deriving ensemble kinematic ages, we can also identify stars that deviate from the typical chemo-dynamical relations, as these stars appear as age outliers, similar to the old low-$\alpha$ stars discussed in the previous section. 
We explore these outliers across different regions of stellar parameter space in \autoref{fig:fig4}, where we show the age--metallicity relation, the [Fe/H]--[$\alpha$/Fe] plane, and the $\ln J_z$--age relation, colored by the average absolute difference between the subgiant ages and the inferred ensemble kinematic ages.
The red ``1'' and dark gray ``2'' mark two regions of parameter space where the subgiant ages are not well reproduced.

The region labeled red ``1'' corresponds to the plateau in true versus inferred ages shown in \autoref{fig:fig3}. 
Although the plateau occurs at $\sim$10~Gyr in ensemble kinematic age, we still achieve an average accuracy of $\lesssim1.5$~Gyr for subgiant ages $\lesssim12.5$~Gyr, which has been identified as the epoch of formation of the rotationally supported stellar disk \citep{Xiang2022}.
This behavior appears largely independent of metallicity, supporting a scenario in which the stellar disk formed rapidly, consistent with previous studies \citep[e.g.,][]{Belokurov2022,Xiang2022}.

Another notable feature appears at ages of $\sim7.5$~Gyr and metallicities between $-1$ and $-0.5$~dex, where ensemble kinematic ages fail to reproduce the subgiant ages (dark gray ``2''). 
This region corresponds to the low-metallicity end of the low-$\alpha$ disk (middle panel of \autoref{fig:fig4}) and to elevated $\ln J_z$ values at $\sim$7.5 Gyr (right panel), suggesting that these stars are dynamically hotter than the rest of the disk population at comparable ages.
This region in age–metallicity space also overlaps with that associated with the Gaia–Sausage–Enceladus \citep[GSE;][]{Belokurov2018, Helmi2018}, raising the possibility that the observed heating is related to the GSE accretion event.
The enhancement in $J_z$ is also visible in the $\ln J_z$--age relation in \autoref{fig:fig2}, where the data points lie above the fitted model. 
Whether these stars were born kinematically hot or subsequently heated by satellite interactions \citep[e.g.,][]{Minchev2009,Bird2013,Lu2024_rb} remains unclear; however, they likely follow a distinct heating history compared to the bulk of the disk at that time. 
Future comparisons with numerical simulations may help clarify their origin.

Finally, we tested the ensemble kinematic age method on subgiant stars, for which \teff\ and \logg\ provide strong constraints on age. 
However, it is important to emphasize that this method requires grouping coeval stars. 
On the lower main sequence (\teff $\lesssim 5200$ K; \citealt{Simonian2019}), \teff\ and \logg\ do not provide sufficient discriminatory power to group stars by age, as stars of different ages significantly overlap in these parameters due to their slow evolutionary timescales. 
As a result, ensemble kinematic ages for lower main-sequence stars provide only statistical constraints on the population as a whole, similar to traditional kinematic age estimates, and cannot be robustly determined without incorporating additional indicators such as rotation periods. 
Furthermore, because co-evolved groups cannot be identified based on CMD position for these stars, the inferred kinematic ages represent a weighted average of the underlying population and are therefore more difficult to interpret.

\section{Ensemble Kinematic Ages for 1.5 Million LAMOST Stars}\label{sec:results}
After validating the method in the previous section, we applied it to infer ensemble kinematic ages for approximately 1.5 million stars with LAMOST DR5 low-resolution spectra, adopting stellar parameters from \citet{Xiang2019}. 
Rather than restricting the analysis to stars with full 6-D phase-space information from Gaia DR3, we extend the method to the full sample.
To do so, for each target star, we construct its ensemble by selecting comparison stars with available 6-D kinematics that occupy similar regions of stellar parameter space. 
In this way, we can assign ensemble kinematic ages even to stars lacking complete kinematic measurements.

We group stars in $\log g$, $T_{\rm eff}$, [Fe/H], [C/Fe], [Mg/Fe], [Al/Fe], [Si/Fe], [Ca/Fe], [Ti/Fe], [Cr/Fe], [Mn/Fe], [Ni/Fe], and [Ba/Fe], with bin sizes of 30~K, 0.05~dex, 0.03~dex, 0.06~dex, 0.08~dex, 0.12~dex, 0.09~dex, 0.05~dex, 0.06~dex, 0.07~dex, 0.10~dex, 0.06~dex, and 0.19~dex, respectively. 
We select elements with the highest reported precision and with fewer than 10\% of measurements flagged as unreliable.
The bin sizes are determined from the mean reported uncertainties of each parameter. 
Because the $\ln J_z$--age relation is empirically calibrated on the subgiant sample (see \autoref{fig:fig2}), we do not extrapolate beyond the calibrated range; consequently, we exclude stars with median $\ln J_z < 0.5$.

\begin{figure*}
\centering
\includegraphics[width=0.8\textwidth]{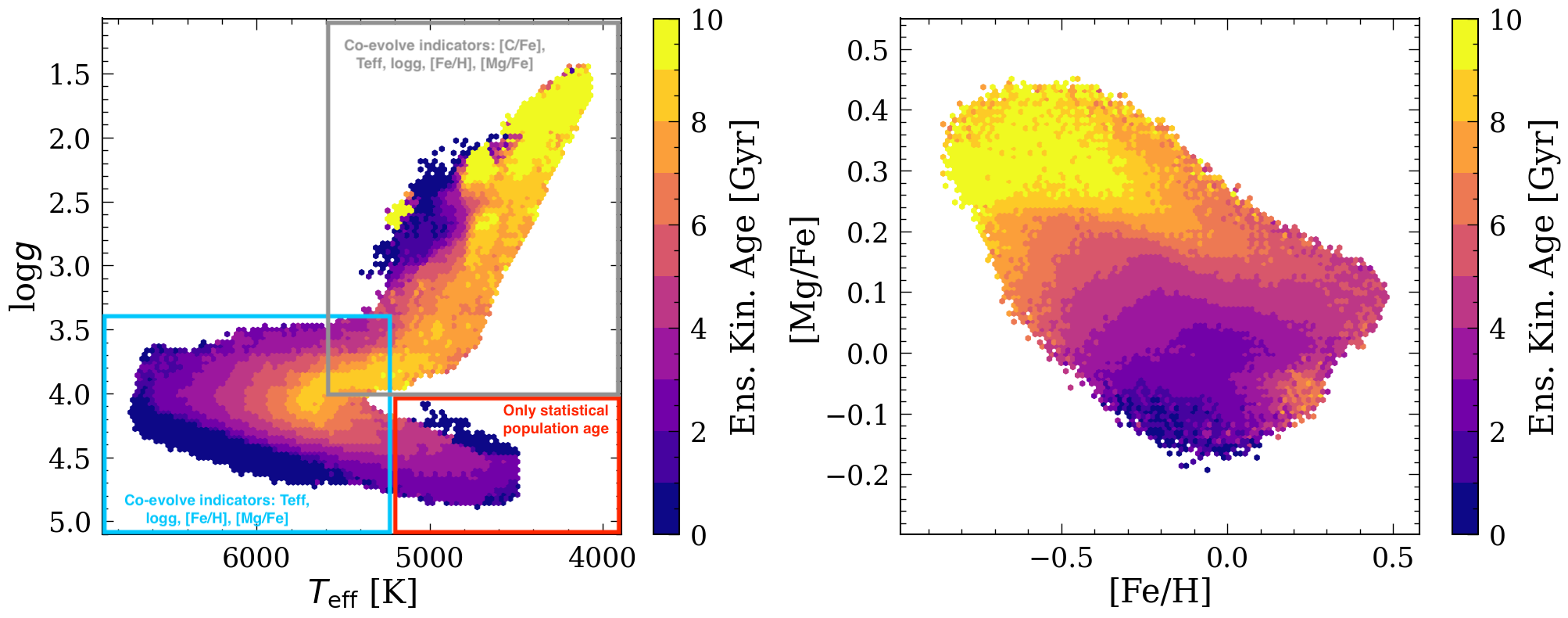}
\includegraphics[width=\textwidth]{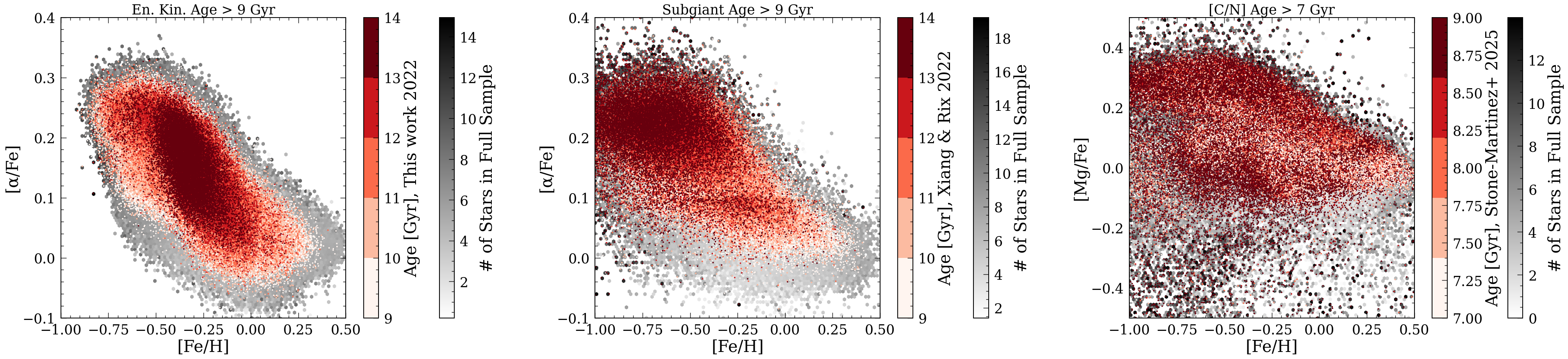}
\caption{Top row: Kiel diagram (\teff-\logg; left) and the [Mg/Fe]--[Fe/H] plane (right), colored by the average ensemble kinematic age. 
The overall gradients are broadly consistent with expectations from stellar evolution and chemical evolution models. 
Boxes in the Kiel diagram indicate the parameters used to define coeval groups for determining ensemble kinematic ages. 
For lower main-sequence stars, however, \teff\ and \logg\ do not provide strong age diagnostics; 
therefore, the resulting ages represent population averages and cannot be reliably assigned to individual stars.}
However, old stars are present at low [Mg/Fe] and relatively high metallicity (bottom-left panel). 
Bottom row: comparison with other age catalogs. 
The background shows the overall [$\alpha$/Fe]--[Fe/H] distribution from LAMOST (middle) and APOGEE (right), while colored points indicate old stars identified using subgiant ages \citep[left;][]{Xiang2022} and [C/N]-based ages \citep[right;][]{stonemartinez2025}. 
If not driven by systematics, these old low-$\alpha$ stars may suggest parallel formation of the high- and low-$\alpha$ disks (see \autoref{A1:diskform}).
\label{fig:fig5}
\end{figure*}

As discussed in \autoref{subsec:suitablespace}, obtaining ensemble kinematic ages is most accurate for stars known to be coeval. 
For subgiants and upper main-sequence stars, \teff, \logg, [Fe/H], and [$\alpha$/Fe] provide strong age constraints, while for giants, [C/Fe], [Fe/H], \teff, and \logg\ serve the same purpose. 
However, for lower main-sequence stars, none of the parameters currently used provide sufficient age discrimination. 
Without additional information such as rotation or activity, ensemble kinematic ages for these stars can only provide population-level constraints, similar to traditional kinematic ages.

\autoref{fig:fig5} shows the Kiel diagram and the [Mg/Fe]--[Fe/H] plane colored by the average ensemble kinematic age.
In the Kiel diagram, the age gradients broadly agree with expectations from isochrones, with younger dwarfs and hotter stars and progressively older giants and cooler stars. 
Similarly, in chemical space, the high-$\alpha$ disk is predominantly old, while the low-$\alpha$ disk is generally younger, consistent with standard Galactic chemical evolution scenarios.

However, a population of old stars is evident at very low [Mg/Fe] ($\sim$-0.1) and high metallicity ($\sim$0.2). 
These old low-$\alpha$ stars (bottom-left panel of \autoref{fig:fig5}) are predominantly giants and are predicted in scenarios where the high- and low-$\alpha$ disks form in parallel \citep[e.g.,][]{Clarke2019}, see also \autoref{A1:diskform}. 
They are also present in independent age catalogs. 
The bottom-middle panel shows stars with subgiant ages $>$ 9~Gyr \citep{Xiang2022}, overlaid on the full LAMOST abundance distribution \citep{Xiang2019}. 
The bottom-right panel shows stars with [C/N]-based ages $>$ 7~Gyr \citep{stonemartinez2025}, overlaid on APOGEE DR17 abundances \citep{abdurrouf2022}. 
If these stars are not the result of systematics, binary evolution, or age uncertainties, they may indicate either parallel disk formation \citep[e.g.,][]{Clarke2019, Beane2025}.
For a more in depth analysis, see \autoref{A1:diskform}. 
Nevertheless, further investigation is required before drawing firm conclusions.

\autoref{tab:1} summarizes the columns of the resulting ensemble kinematic age catalog for the $\sim$1.5 million LAMOST DR5 stars.

\begin{table}
\centering
\caption{Column descriptions for the ensemble kinematic age catalog of $\sim$1.5 million stars. The catalog is available on Zenodo (\url{https://zenodo.org/records/21178926}).
We provide the median vertical action to enable readers to construct their own age–action relation.}
\begin{tabular}{ccc}
\hline
\hline
Column & Unit & Description \\
\hline
\texttt{STARID} &  & LAMOST star identifier \\
\texttt{RA} & deg & Right ascension from LAMOST \\
\texttt{DEC} & deg & Declination from LAMOST \\
\texttt{Jz\_med} & kpc\,km\,s$^{-1}$ & Inferred median vertical action \\
\texttt{Jz\_age} & Gyr & Inferred ensemble kinematic age \\
\texttt{TEFF} & K & $T_{\rm eff}$ from \citet{Xiang2019} \\
\texttt{LOGG} & dex & $\log g$ from \citet{Xiang2019} \\
\hline
\end{tabular}
\label{tab:1}
\end{table}

\begin{figure*}
\includegraphics[width=0.35\textwidth]{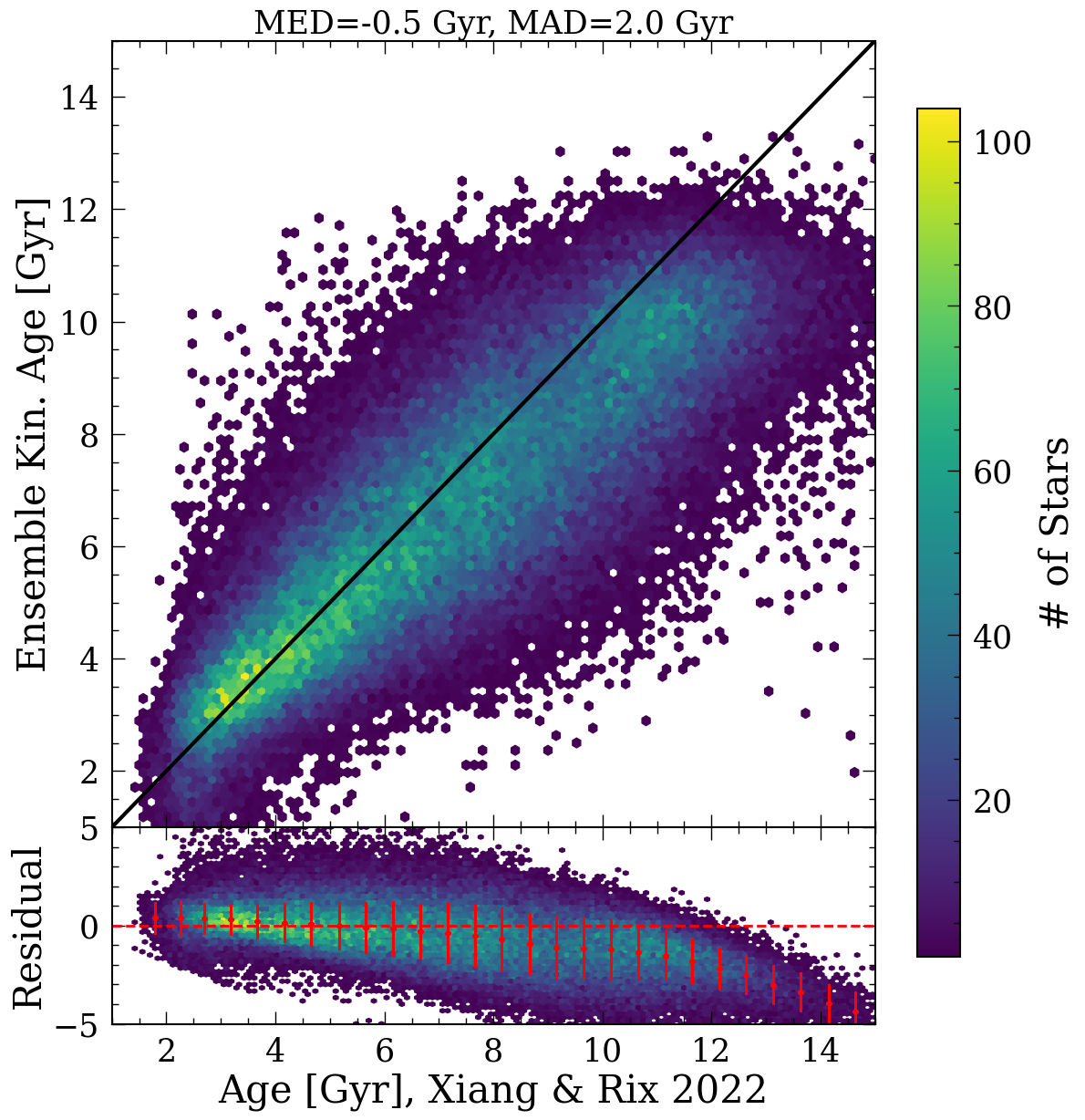}
\includegraphics[width=0.35\textwidth]{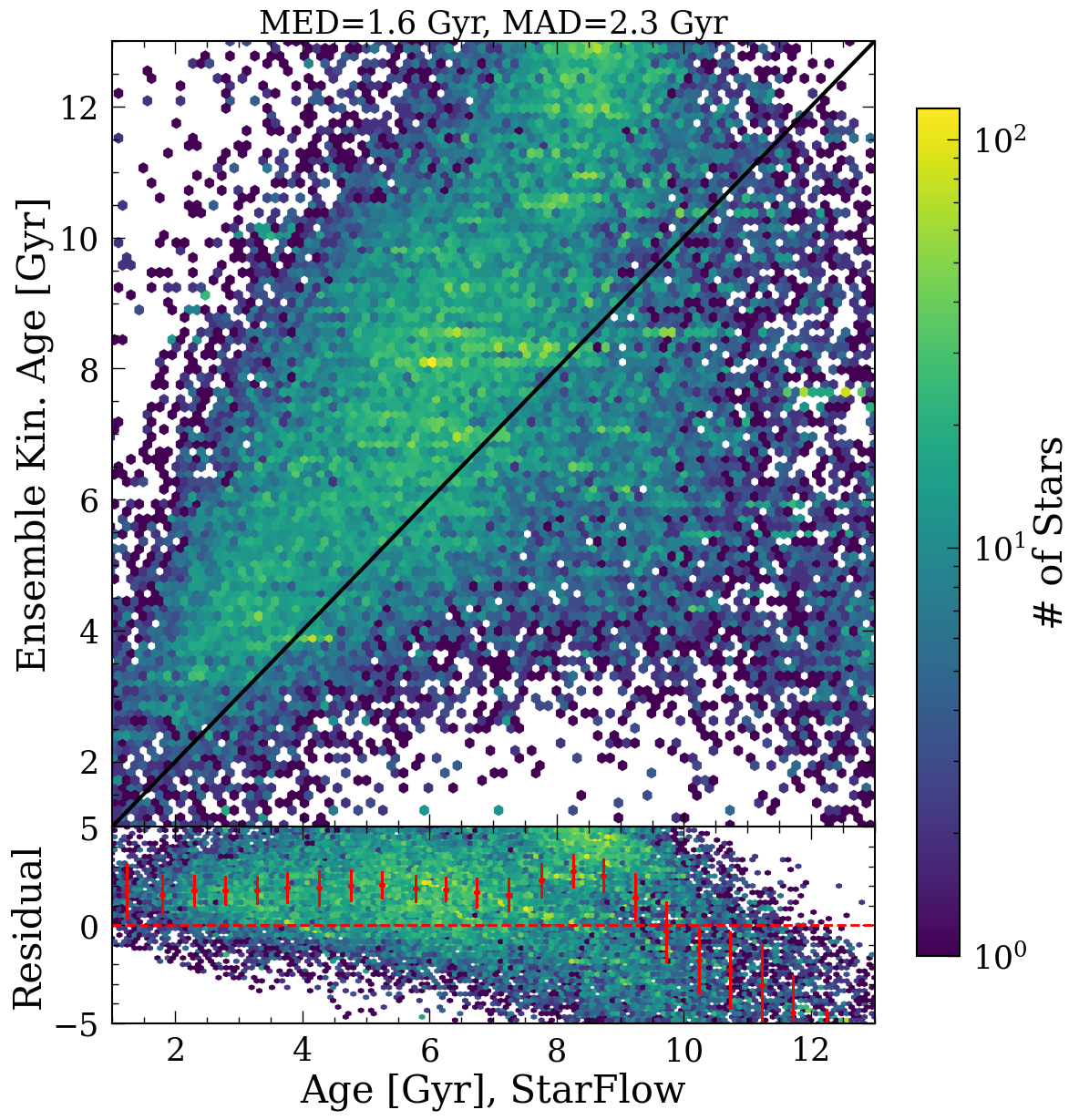}
\includegraphics[width=0.3\textwidth]{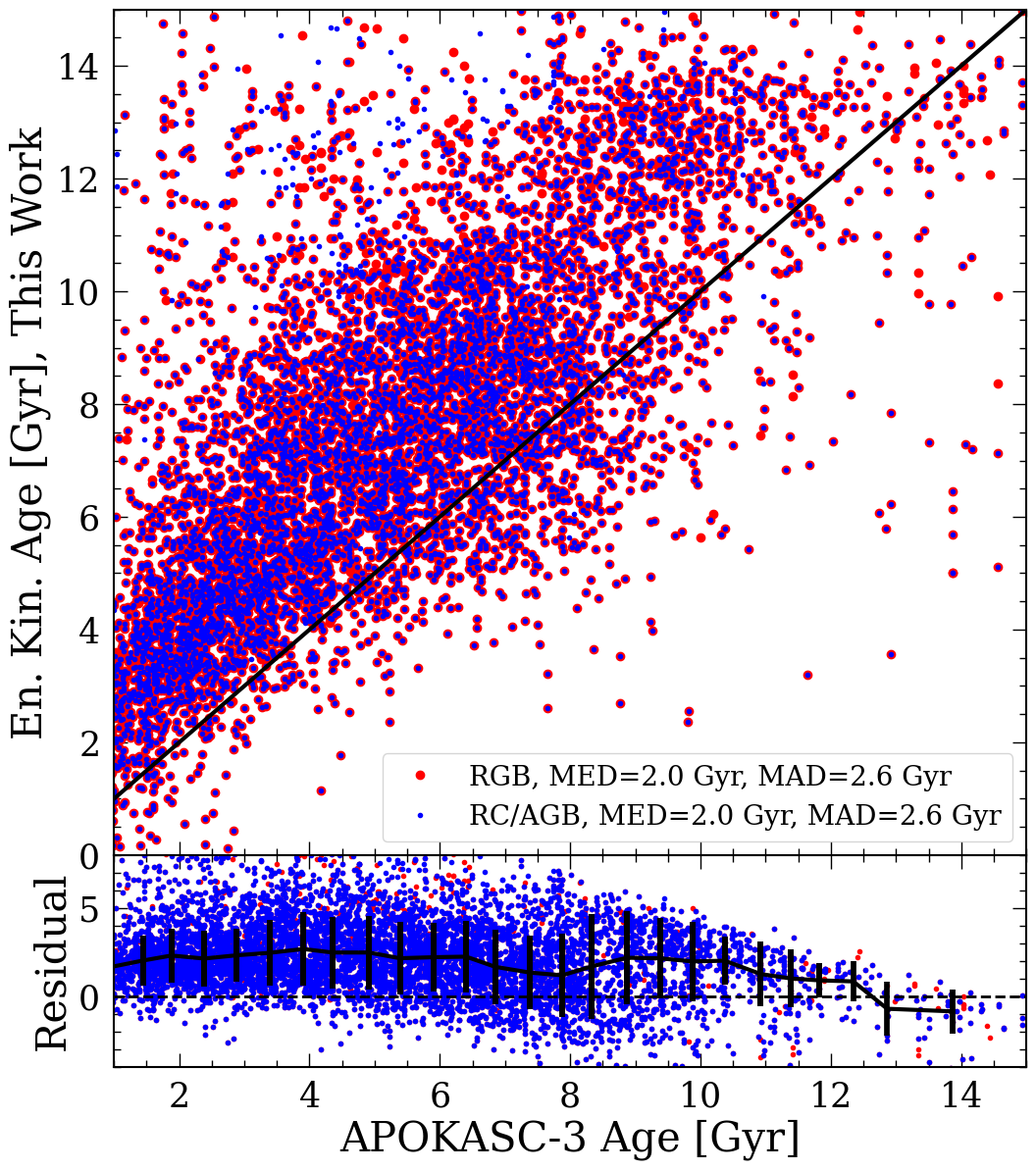}
\includegraphics[width=0.51\textwidth]{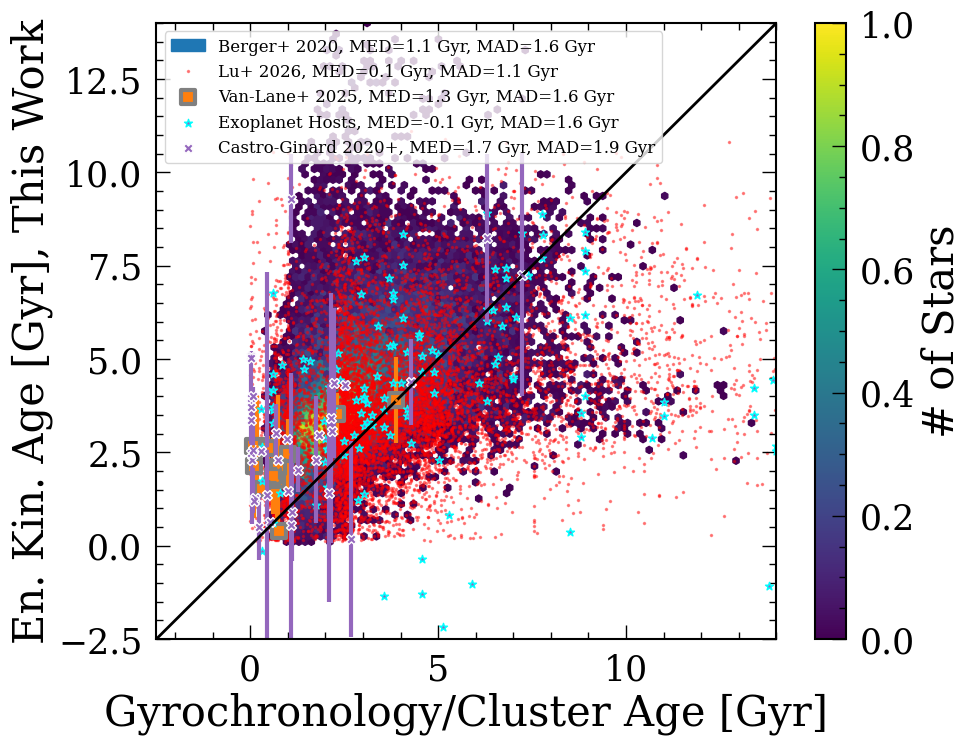}
\includegraphics[width=0.53\textwidth]{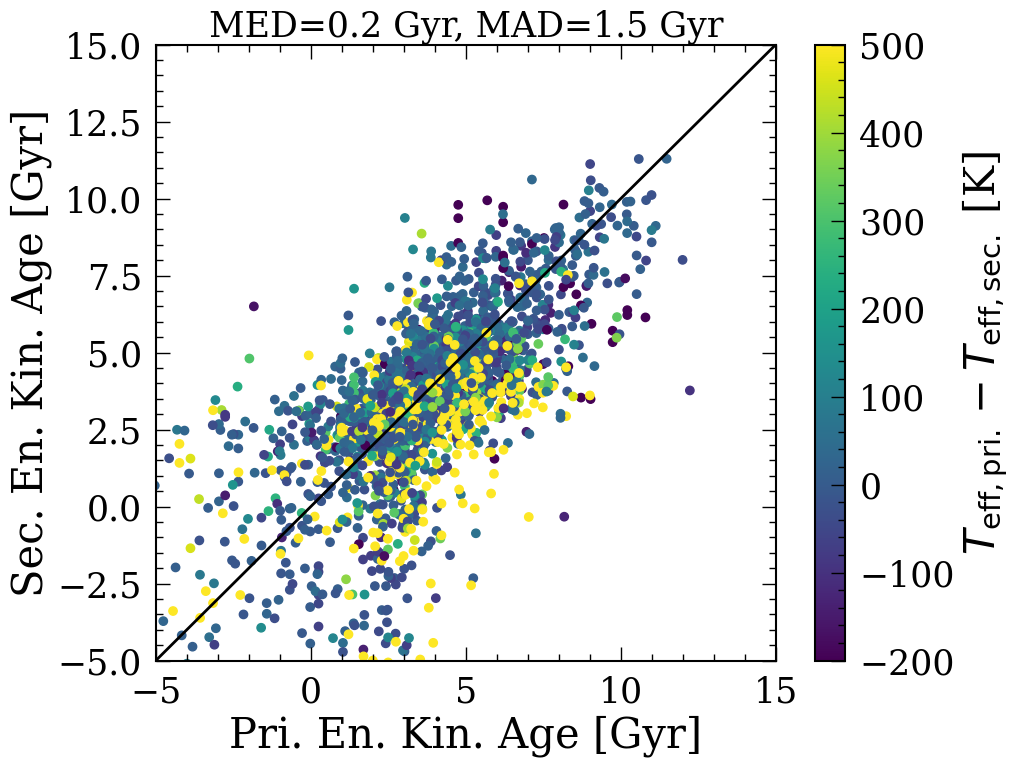}
\caption{Comparison of ensemble kinematic ages with independent age estimates: subgiant ages \citep[top left;][]{Xiang2022}, [C/N] ages from \texttt{starFlow} \citep[top middle;][]{stonemartinez2025}, APOKASC--3 asteroseismic ages \citep[top right;][]{pinsonneault2024}, main-sequence ages from isochrone fitting and gyrochronology, and open-cluster ages \citep[bottom left;][]{Berger2020,Lu2024,CastroGinard2020,vanlane2025}. 
The bottom-right panel compares ages of wide binaries \citep{elbadry2021}, primarily main-sequence stars. 
On average, ensemble kinematic ages show a consistent MAD of $\sim$2~Gyr ($\sim$30\%), agreeing with the validation tests in \autoref{sec:kinage}.}
\label{fig:fig6}
\end{figure*}

\subsection{Comparison with other literature ages}
To further validate our ages, we compare our ensemble kinematic ages with subgiant ages from \citet{Xiang2022}, spectroscopic ages from \cite{stonemartinez2025}, asteroseismic ages for giants from APOKASC--3 \citep{pinsonneault2024}, gyrochronology ages for main-sequence stars from \citet{Lu2024}, and isochrone ages for main-sequence stars from \citet{Berger2020} and the NASA Exoplanet Archive \citep{Christiansen2025}.\footnote{\url{https://exoplanetarchive.ipac.caltech.edu}, accessed 2026 January 29.}
We also include open cluster ages from \citet{vanlane2025} and \citet{CastroGinard2020}, as well as wide-binary pairs, which are predominantly main-sequence stars, from \citet{elbadry2021}.
For clusters, we assign ages by first inferring ensemble kinematic ages for individual member stars and then computing the mean and standard deviation as the cluster age and its uncertainty, respectively.
Overall, we are able to infer ages across a wide range of stellar evolutionary stages with a consistent median absolute deviation (MAD) of $\sim$2~Gyr, corresponding to $\sim$30\% in relative uncertainty.
This level of agreement is consistent with our tests on wide binaries (\autoref{fig:fig6}, bottom right) and on the subgiant calibration sample alone (see \autoref{fig:fig3} and \autoref{sec:kinage}).
The median offset between our ages and those from different catalogs ranges from $\sim$0.5 to 2~Gyr.
Such differences are expected, as each age-dating method relies on different stellar evolution models, and abundance scales from different surveys (e.g., LAMOST for the subgiant sample and APOGEE for APOKASC--3) may introduce systematic offsets.
Not surprisingly, our ensemble kinematic ages show the smallest bias relative to the subgiant ages, since the $\ln J_z$--age relation is calibrated on this sample.
We find that our method overpredicts APOKASC--3 ages by $\sim$2~Gyr on average, which may indicate a systematic offset between the subgiant and APOKASC--3 age scales.
More broadly, this comparison highlights the potential of ensemble kinematic ages to serve as a cross-calibration tool among different age-dating techniques, even when they are applied to stars at different evolutionary stages.

\subsection{Potential science cases}
In this section, we present several applications of our age estimates in both stellar physics (\autoref{subsubsec:empIso}) and Galactic archaeology (\autoref{subsubsec:galactic_arch}). 
Beyond these examples, self-consistent age estimates across the entire HR diagram enable a wide range of population studies, including investigations of exoplanet demographics across different stellar evolutionary stages, allowing planet-hosting stars to be analyzed within a uniform age framework.

\subsubsection{Ensemble kinematic ages for stellar physics}\label{subsubsec:empIso}
\begin{figure}
    \centering
    \includegraphics[width=\linewidth]{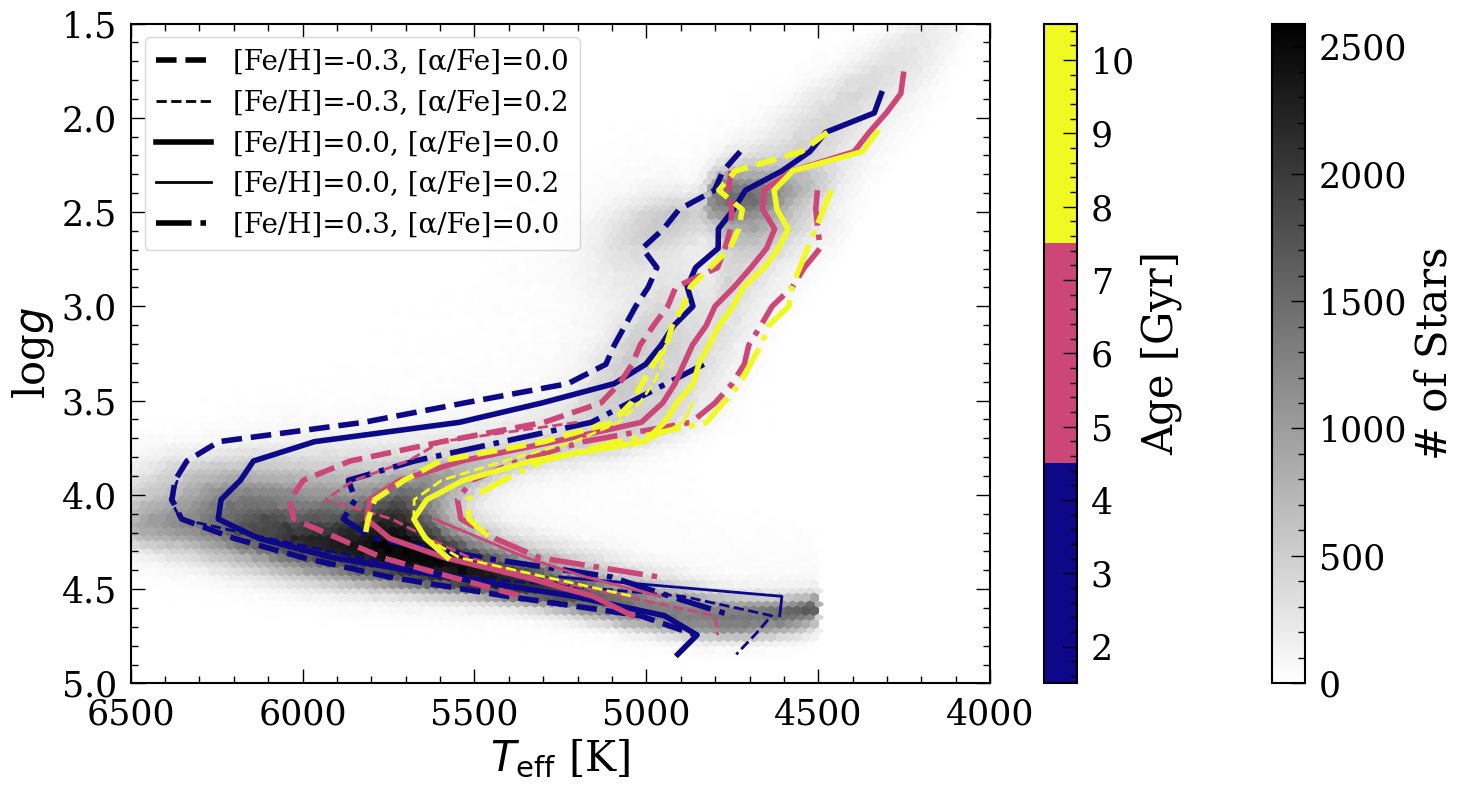}
    \caption{Empirical isochrones constructed with our ensemble kinematic ages for stars with age = 3, 6, or 9 Gyr, [Fe/H] = $-$0.3, 0, or 0.3, and [$\alpha$/Fe] = 0 or 0.2 as indicated by the color and legends.
    As expected, at the same age, metal-poor stars are hotter, and the effect of [$\alpha$/Fe] increases towards higher metallicity, agreeing with stellar evolution models \cite{Dotter2026}.
    These empirical isochrones provide a largely model-independent benchmark for testing stellar evolution calculations, with the caveat that the age-$\ln J_z$ relation is calibrated with ages that are model dependent.
    This allow the effects of metallicity and individual elemental abundances on stellar structure, atmospheric opacity, and evolutionary tracks to be directly constrained from observations. 
    They also offer a valuable calibration dataset for improving theoretical isochrones and quantifying systematic uncertainties in stellar models.}
    \label{fig:iso}
\end{figure}

With our large age catalog, we are able to construct empirical isochrones to understand stellar physics.
To do so, we select stars within 0.5 Gyr of the target ages (3, 6, or 9 Gyr), within 0.07 dex of the target metallicities ([Fe/H] = $-$0.3, 0, or 0.3), and within 0.08 dex of the target [$\alpha$/Fe] values (0 or 0.2). 
We then compute the median \teff\ in \logg\ bins spanning between 1 and 5. 
Only bins with more than 50 stars are included.
Figure~\ref{fig:iso} shows the resulting empirical isochrones.

As expected, at fixed age, metal-poor stars are hotter, and the effect of [$\alpha$/Fe] becomes more pronounced at higher metallicity, consistent with stellar evolution models \cite{Dotter2026}.
This demonstrates the potential for empirically calibrating stellar evolution models across a wide range of ages, metallicities, and [$\alpha$/Fe].
However, it is important to note that these empirical isochrones are subject to the chemical evolution history of the Galaxy. 
For example, young, high-$\alpha$ stars are rare and likely do not represent the underlying stellar population, as they may arise from special formation pathways or binary evolution.

\begin{figure*}
    \centering
    \includegraphics[width=\linewidth]{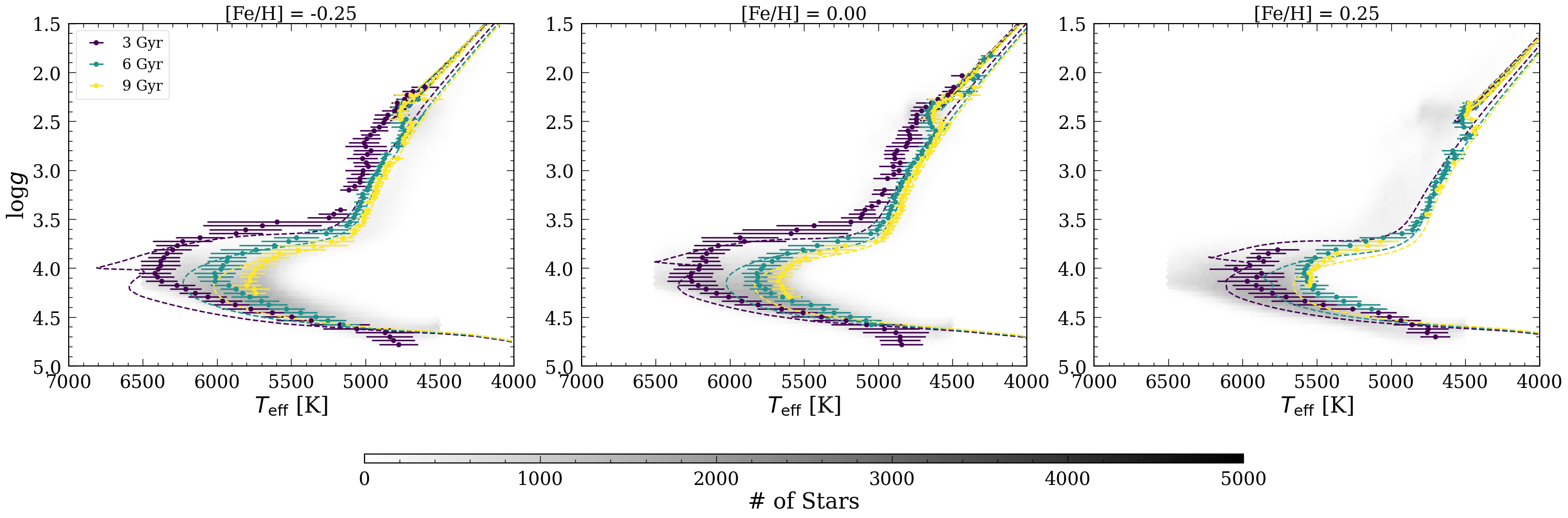}
    \caption{Comparison between empirical isochrones (points) and MIST isochrones (dashed lines) for stellar populations with ages of 3, 6, and 9 Gyr, [Fe/H] = -0.25, 0.0, +0.25, and [$\alpha$/Fe] = 0. 
    The errorbars are the 1.5*median absolute deviation of the median \teff. 
    The empirical isochrones look topologically similar to the theory isochrones.
    An offset exist between the empirical and theory isochrones  for the main-sequence stars. 
    These comparisons illustrate the ensemble kinematic ages can successfully infer stellar ages, in agreement with theoretical expectations, even for populations with similar chemical compositions.
    They also demonstrate the potential for using empirical constraints to calibrate theoretical isochrones.}
    \label{fig:iso_theory}
\end{figure*}

\autoref{fig:iso_theory} presents a comparison between empirical isochrones (shown as points) at solar [$\alpha$/Fe] and MIST isochrones \citep{Dotter2016, Choi2016}. 
The empirical isochrones are constructed in the same manner as in \autoref{fig:iso}, but using finer $\log g$ bins to enable a more detailed comparison.
The empirical isochrones look topologically similar to the theory isochrones.
An offset exists between the empirical and theoretical isochrones along the main sequence. 
This discrepancy may arise from differences in the temperature or $\log g$ scales between the observations and the models, a systematic age offset, or underlying population differences between the subgiants that are used to calibrate the age–$J_z$ relation and the main-sequence stars.
Despite these discrepancies, the overall agreement between the models and the data, after applying these adjustments, is reasonably good. This highlights the potential of using empirical constraints to calibrate theoretical isochrones.

\subsubsection{Ensemble kinematic ages for Galactic Archaeology}\label{subsubsec:galactic_arch}
With self-consistent ages across the HR diagram, one can model the chemo-dynamical history of the MW with reduced bias beyond the survey selection function alone.
Standard chemical evolution studies often calibrate or validate against giant stars using [C/N]-based ages. 
However, restricting the sample to giants introduces additional, non-trivial systematics: giants occupy a narrow and evolution-dependent region of parameter space, and their surface abundances are altered by dredge-up and mixing processes (e.g., first dredge-up), which makes [C/N] a mass- and evolutionary-state-dependent age proxy rather than a purely initial-composition tracer. 
As a result, both evolutionary and abundance–evolution coupling biases compound the usual selection effects, potentially skewing inferred age–metallicity relations and enrichment histories.

\begin{figure}
    \centering
    \includegraphics[width=\linewidth]{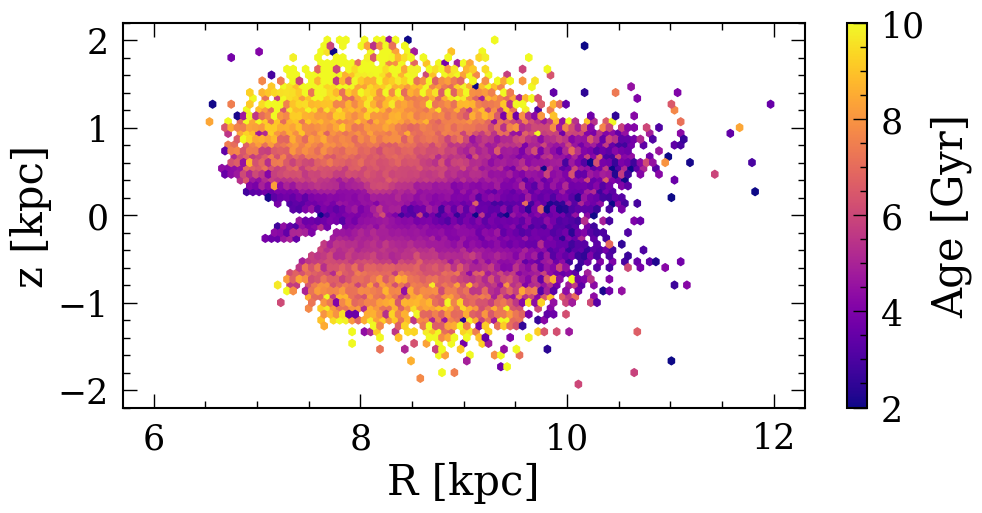}
    \includegraphics[width=\linewidth]{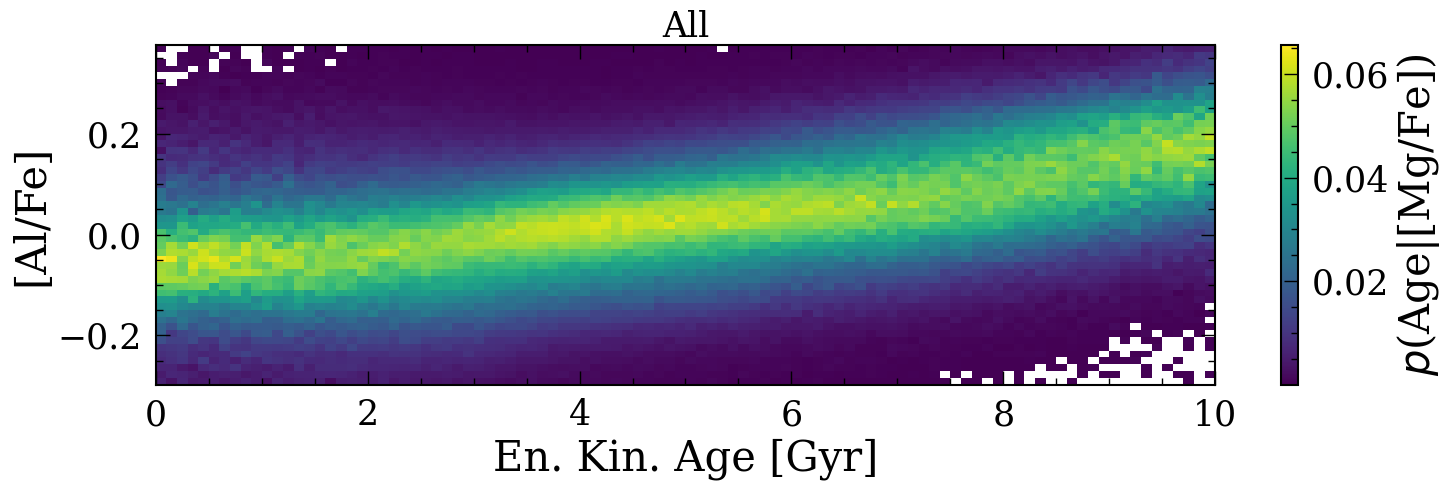}
    \includegraphics[width=\linewidth]{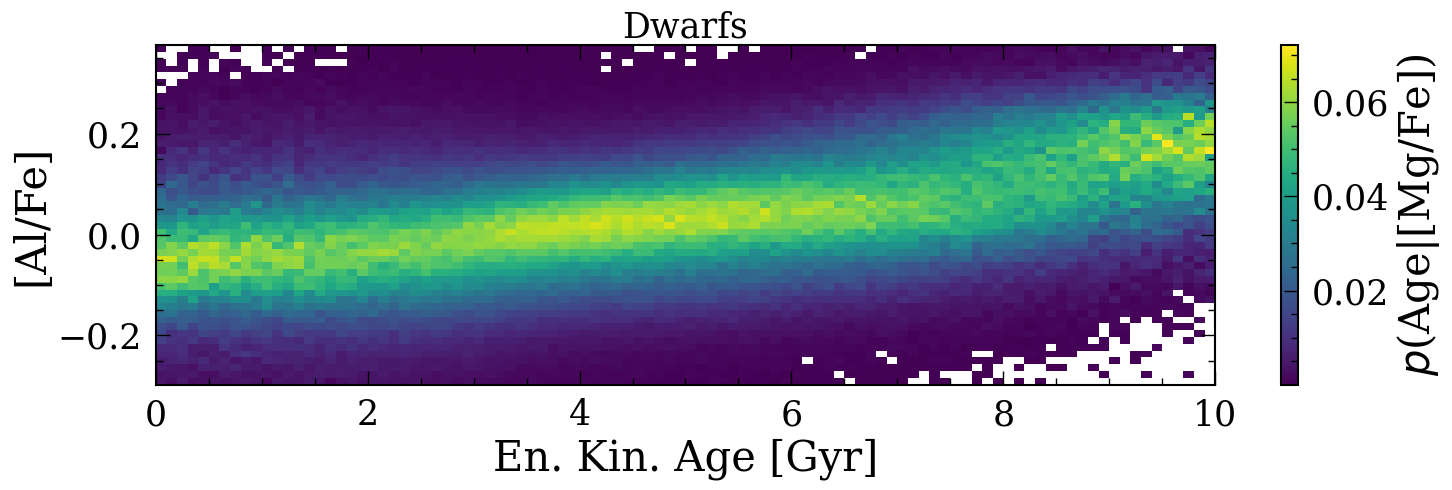}
    \includegraphics[width=\linewidth]{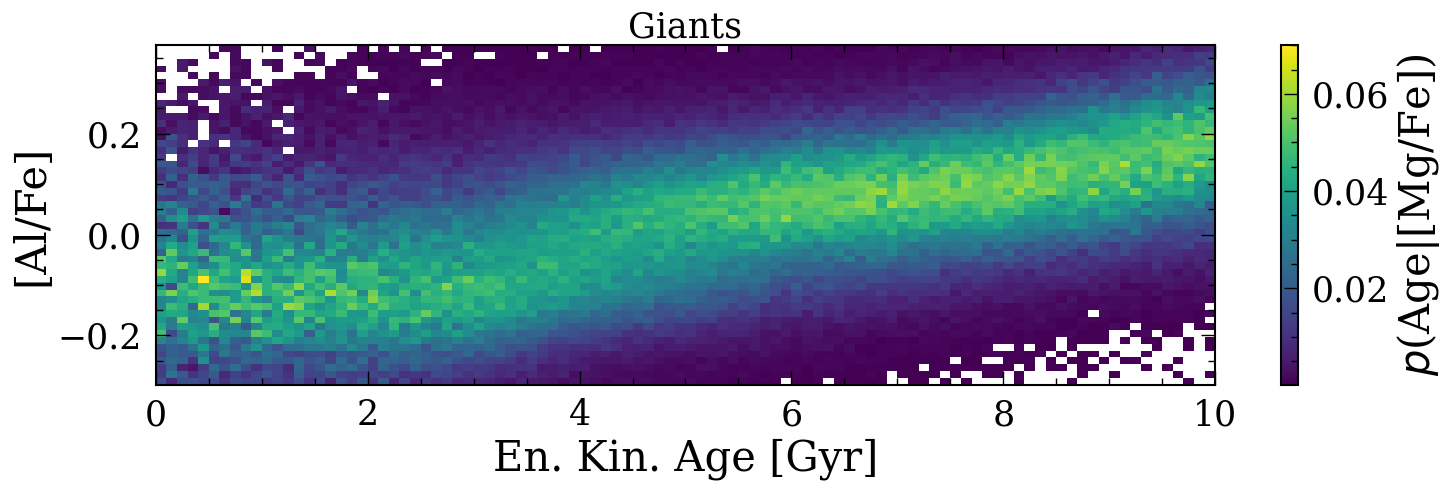}
    \caption{
    Top: Median stellar age in the Galactic $R$--$z$ plane, showing the expected age stratification, with the thin disk dominated by younger stars and progressively older populations at larger distances from the Galactic plane in the thick disk. 
    Bottom: Column normalized age--[Al/Fe] relations for the full sample (left), dwarfs ($\log g > 4$; center), and giants ($\log g < 4$; right). 
    In all samples, [Al/Fe] decreases with decreasing age, consistent with Galactic chemical evolution. 
    The similar trends observed for both dwarfs and giants demonstrate that the derived ages are self-consistent across different evolutionary stages.
    The slight difference may indicate additional biases introduced by selecting stars in specific evolutionary stages.}
    \label{fig:chemo}
\end{figure}

\autoref{fig:chemo} showcases the derived stellar ages. 
The top panel presents the age distribution in the $R$--$z$ plane, while the bottom three panels show the age--[Al/Fe] relation for the full sample, dwarfs ($\log g > 4$), and giants ($\log g < 4$), as indicated in each panel. 
The spatial age distribution follows the expected Galactic structure where the thin disk is dominated by younger stars, and stellar ages increase with distance from the Galactic plane into the thick disk. 
Likewise, [Al/Fe] decreases toward younger ages, consistent with the expected chemical enrichment history.

\section{Limitations}\label{sec:limitation}
\begin{figure*}
    \centering
    \includegraphics[width=\linewidth]{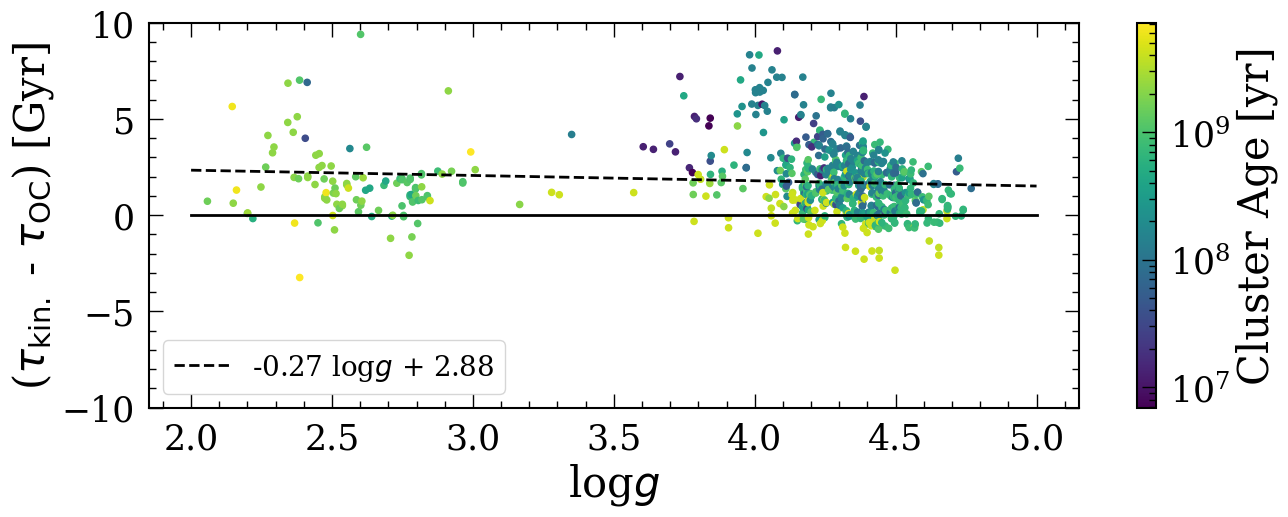}
    \caption{The difference in ensemble kinematic ages and open cluster ages as a function of \logg, colored by the cluster age.
    The ensemble kinematic ages are inferred for individual cluster members.
    The dashed line show the linear fit between the residual and \logg.
    There is no strong correlation between the difference in age and \logg, meaning we can infer ages consistently across the HR diagram.}
    \label{fig:7}
\end{figure*}

Although our ensemble kinematic ages generally agree with literature values with comparable scatter, it remains unclear whether dwarf and giant stars are placed on exactly the same absolute age scale.
One concern is that [C/Fe] is likely a stronger age indicator in giants than in dwarfs due to the first dredge-up \citep{Iben1967}. 
This means it is likely that our assumption 1 mentioned in \autoref{subsec:method} is better defined in giants compared to dwarfs.
In addition, stellar parameter determinations, including \teff, \logg, and elemental abundances derived from spectral synthesis, may exhibit systematic differences between dwarfs and giants, potentially arising from effects such as magnetic activity \citep[e.g.,][]{Cao2022}, which may violate assumption 2.
Nevertheless, examining \logg\ as a function of the age difference between ensemble kinematic and open cluster ages (\autoref{fig:7}) reveals no strong dependence on \logg, suggesting that our method yields broadly consistent ages across evolutionary stages.
However, since most open clusters have near-solar metallicity, we are unable to robustly test this consistency in the metal-poor regime.

Furthermore, ensemble kinematic ages, like all age-dating methods, are reliable only for stars that follow the typical chemo-dynamical relations of the Galactic disk.
They are therefore not expected to apply to accreted populations, binary products, or merger remnants, as those stars do not follow assumption 4.
For stars identified as age outliers (e.g., numbered parameter space in \autoref{fig:fig5}), it is difficult to determine whether they are genuinely anomalous in age or instead atypical in their chemo-dynamical properties.
Even so, detailed investigations of stars whose ensemble kinematic ages disagree with other age estimates may provide valuable insight into disk heating, accretion events, and the broader processes governing Galactic formation and evolution.

Finally, and perhaps more importantly, ensemble kinematic ages are most reliable for stars known to be coeval, as mentioned in assumption 1. 
Consequently, additional age indicators beyond kinematics are needed to properly group stars in order to achieve more accurate ages than those obtained from traditional kinematic age-dating methods.
Since co-evolved groups cannot be identified based on CMD position for these stars, the inferred kinematic ages represent a weighted average of the underlying population and are also more difficult to interpret.
However, ensemble kinematic ages can serve as a powerful tool for calibrating gyrochronology \citep[e.g.,][]{Lu2021_kinage, Lu2026, Lu2024_Mdwarf}. 
Since stellar rotation periods are well-established age indicators, and the majority of dwarf stars reside in the thin disk with ages $<\sim$ 10 Gyr, where this method is applicable, this approach provides a robust framework for cross-calibrating stellar age estimates.

\section{Conclusion} \label{sec:con}
Gaia DR3, and the forthcoming DR4, now provide 6-D kinematic information for millions of Milky Way stars, enabling stellar kinematics to be used as a powerful statistical age indicator.
Although kinematic ages for individual stars are intrinsically noisy, we demonstrate that meaningful and precise age information emerges when kinematics are averaged over large stellar populations with similar atmospheric and chemical properties.
By identifying co-evolve stars in stellar-parameter space and measuring their mean vertical action, \jz, we define ensemble kinematic ages that are internally self-consistent across the Kiel diagram and chemical-abundance planes.
We apply this method to $\sim$1.5 million LAMOST DR5 stars using atmospheric parameters and detailed elemental abundances.
This approach yields ages for both dwarfs and giants with a typical uncertainty of $\sim$2~Gyr, comparable to the precision achieved by [C/N]-based ages.
The resulting ensemble kinematic ages show good agreement with independent constraints from subgiants, asteroseismology (APOKASC--3), open clusters, gyrochronology, isochrone fitting, and wide-binary systems.
In particular, comparisons with individual open-cluster members reveal no strong systematic trend with \logg, suggesting that dwarf and giant stars are placed on a consistent age scale.
These results demonstrate that kinematic ages are not inherently unreliable; rather, their utility depends on treating them as population-level measurements.
When used in this statistical sense, kinematics provide robust average ages, following the same principles that underpin other widely adopted stellar age indicators.
At the same time, ensemble kinematic ages are intrinsically statistical and are applicable only to stars that follow the typical chemo-dynamical relations of the Galactic disk.
They are not designed to yield reliable ages for accreted populations, merger products, or other stars with atypical dynamical histories.
Nevertheless, such deviations can themselves be astrophysically informative.
Overall, ensemble kinematic ages provide a robust, complementary age scale for Galactic archaeology in the Gaia era, enabling age information to be extended to large spectroscopic samples where traditional methods are limited or unavailable.

\begin{acknowledgments}
We thank the helpful suggestions from Jos\'e G. Fern\'andez-Trincado and the SDSS-V Galactic Genesis working group.
We thank the anonymous referee for their helpful suggestions.
This work has made use of data from the European Space Agency (ESA) mission Gaia,\footnote{\url{https://www.cosmos.esa.int/gaia}} processed by the Gaia Data Processing and Analysis Consortium (DPAC).\footnote{\url{https://www.cosmos.esa.int/web/gaia/dpac/consortium}} 
Funding for the DPAC has been provided by national institutions, in particular the institutions participating in the Gaia Multilateral Agreement.
This research also made use of public auxiliary data provided by ESA/Gaia/DPAC/CU5 and prepared by Carine Babusiaux. 
% SIMBAD, Vizier, ADS
This research has also made use of NASA's Astrophysics Data System.

\end{acknowledgments}


\vspace{5mm}
\facilities{Gaia \citep{gaia, Gaiadr3}, LAMOST \citep{lamost}, APOGEE \citep{apogee}, Kepler \citep{kepler}, PO:1.2m \citep[ZTF][]{ztfdata, ztftime}, TESS \citep{TESS}, 2MASS \citep{2mass}}


\software{\texttt{astropy} \citep{astropy:2013, astropy:2018, astropy2022}, \texttt{Matplotlib} \citep{matplotlib}, \texttt{NumPy} \citep{Numpy}, \texttt{Pandas} \citep{pandas}, \texttt{SciPy} \citep{SciPy}, \texttt{gala} \citep{gala2017}, \texttt{galpy} \citep{galpy}, \texttt{ChatGPT} \citep{openai2025chatgpt}}



\appendix

\section{Formation of the low- and high-$\alpha$ disk}\label{A1:diskform}
\begin{figure*}
    \includegraphics[width=\textwidth]{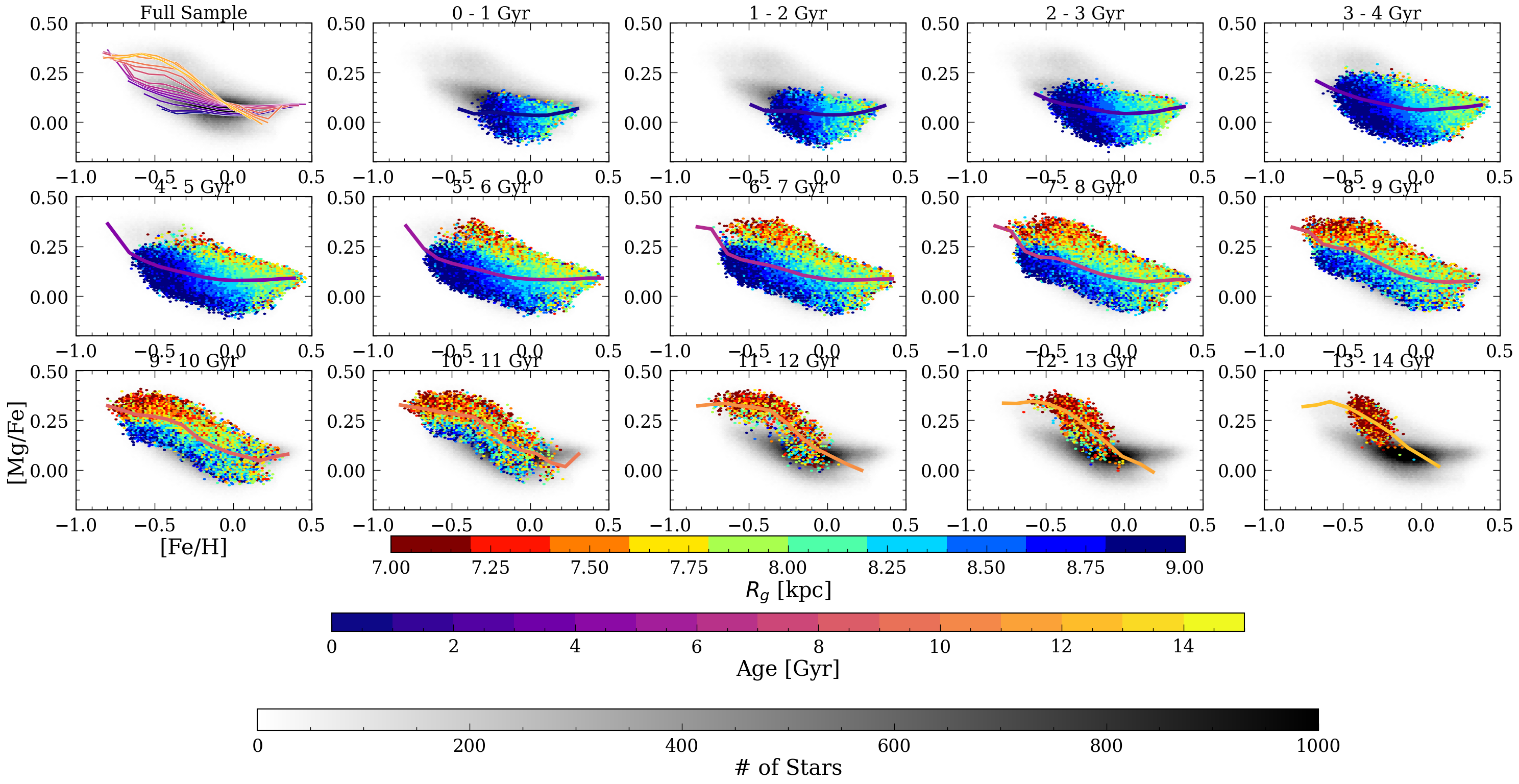}
    \caption{[Mg/Fe]--[Fe/H] plane for the full sample (top left), and in ensemble kinematic age bins of 1~Gyr colored by their average guiding radii.
    The age bins are indicated in the titles.
    The colored lines show the running median for each age bin.
    A large population of old stars ($>$7~Gyr) exists in the parameter space of the low-$\alpha$ disk, which is typically associated with younger populations.}
    \label{fig:A1}
\end{figure*}

We further investigate the age distributions of the low- and high-$\alpha$ disks. 
\autoref{fig:A1} displays the [Mg/Fe]--[Fe/H] plane for the full sample (top left), partitioned into 1~Gyr ensemble kinematic age bins and colored by their average guiding radii. 
The colored lines represent the running median for stars within each bin, as indicated by the color bar.

Notably, a substantial population of old stars ($>$7~Gyr) occupies the low-$\alpha$ disk parameter space, which is typically associated with younger populations. 
A transition between the characteristic high- and low-$\alpha$ disks occurs around 8--9~Gyr, where both populations coexist. 
Conversely, a ``young high-$\alpha$'' population is evident between 4--7~Gyr. 
Initially identified in asteroseismic catalogs, these stars have since been observed in subgiant and main-sequence age samples. 
While many of these stars are likely merger products, current research cannot entirely rule out the existence of genuinely young high-$\alpha$ stars \citep{Chiappini2015, Martig2015, SilvaAguirre2018, Claytor2020, Das2020, Zinn2022, Grisoni2024, Lu2025, Lu2025b}. 
The candidates least likely to be merger products are typically found at [Fe/H]$\sim -0.5$~dex with ages of 4--5~Gyr. 
Our age catalog suggests that such stars may form naturally during the declining phase of high-$\alpha$ disk formation, persisting even after the peak epoch of activity has subsided (see \autoref{fig:A4}, top left). 
Notably, this young population also manifests as a ``young tail'' in other recent age catalogs \citep{Xiang2022, stonemartinez2025}. 
However, caution is warranted, as observational uncertainties and age-dating systematics cannot be entirely excluded.

\begin{figure*}
    \includegraphics[width=\textwidth]{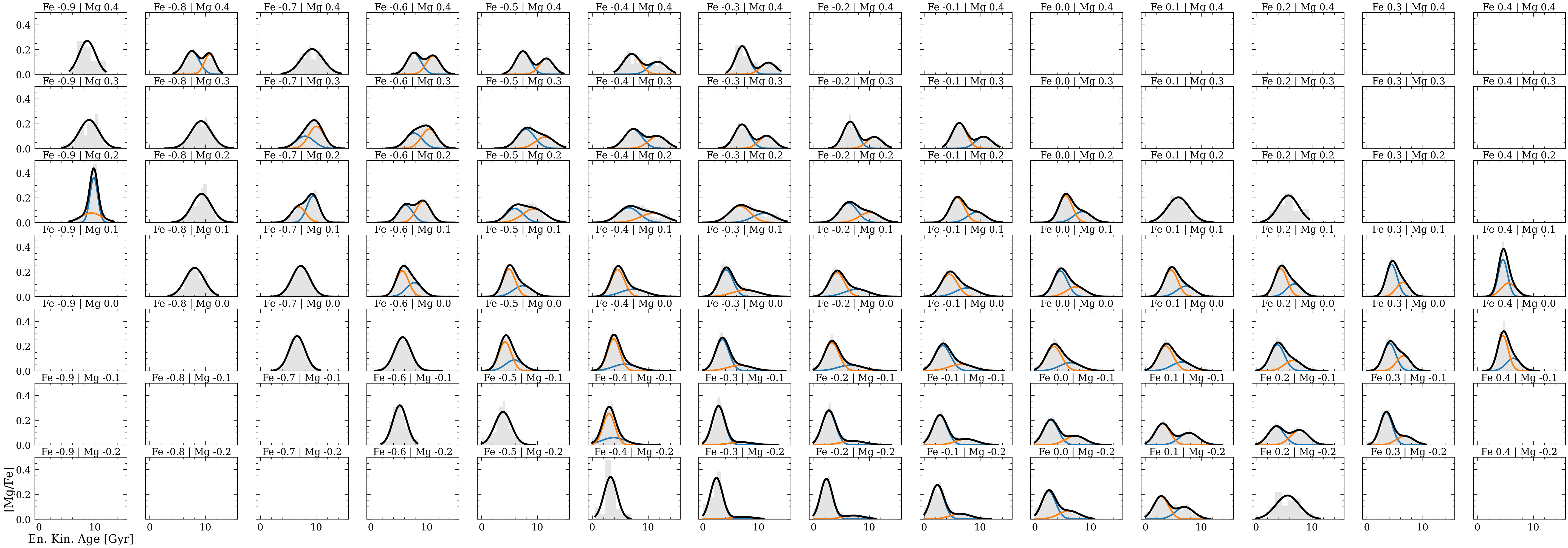}
    \caption{Gaussian mixture modeling (GMM) of the ensemble kinematic ages in bins of [Fe/H] and [Mg/Fe]. 
    The bin widths are 0.1~dex and the titles indicate the bin centers. 
    The number of Gaussian components is determined using the Bayesian information criterion (BIC).
    The black curves show the composite distributions, and the colored histograms show the decomposed components.
    For most bins where the high- and low-$\alpha$ disks overlap (see \autoref{fig:fig5}, right panel), the age distributions are bimodal.
    This suggests that the high- and low-$\alpha$ disks evolve through similar regions of [Mg/Fe]--[Fe/H] parameter space at different times.}
    \label{fig:A2}
\end{figure*}

\begin{figure*}
    \includegraphics[width=\textwidth]{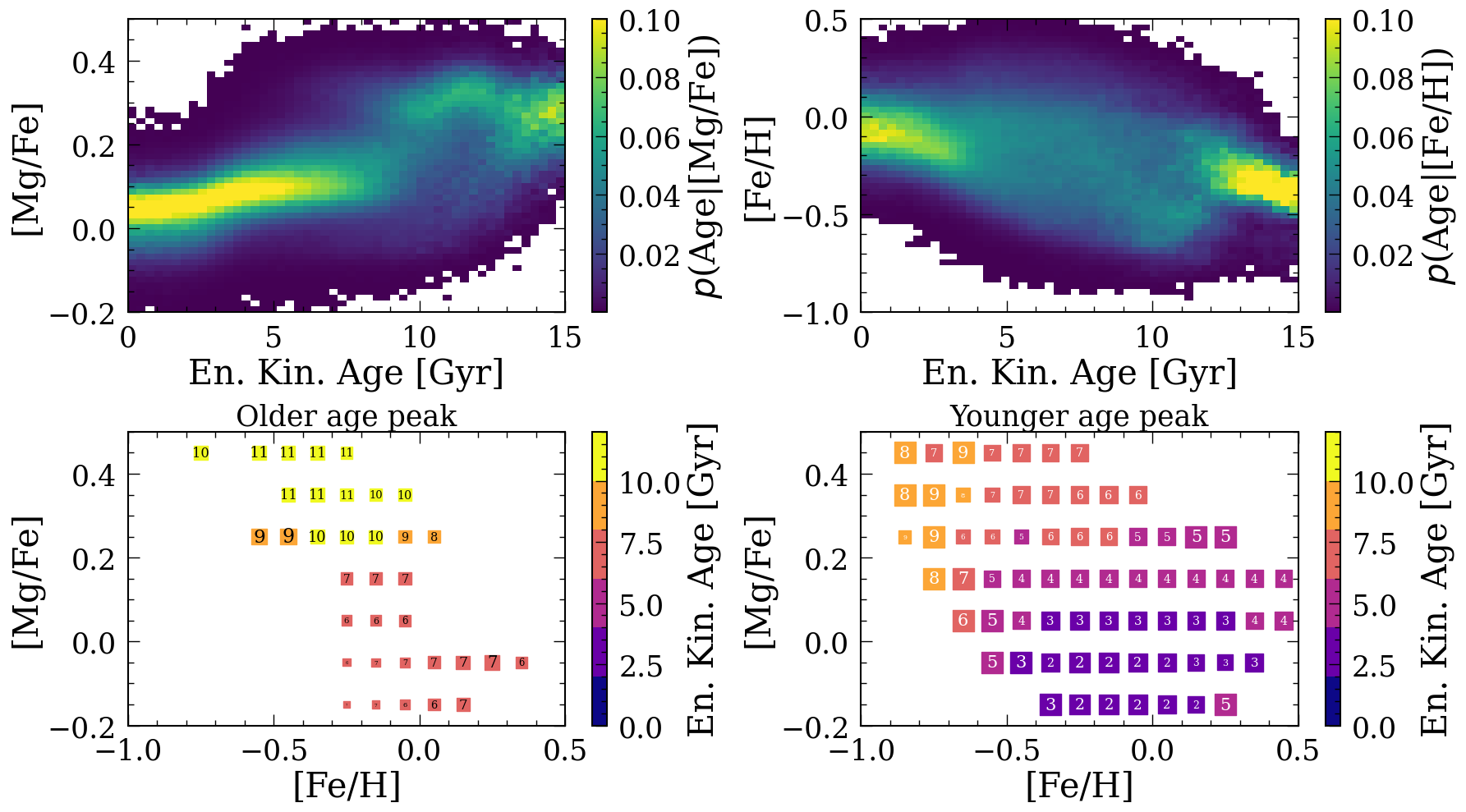}
    \caption{Top row: column-normalized age--[Mg/Fe] (left) and age--[Fe/H] (right) relations.
    Bottom row: peak locations of the Gaussian mixture model (GMM) components for the ensemble kinematic age distributions in the [Mg/Fe]--[Fe/H] plane (see \autoref{fig:A2}). 
    The left panel shows peaks of the older population, while the right panel shows peaks of the younger population. 
    The sizes of the squares indicate the relative weights of each population, and the numbers denote the peak ages.
    The bimodal age distribution in the [Mg/Fe]--[Fe/H] plane corresponds to the two populations seen in the age--[Mg/Fe] relation at ages $> 8$~Gyr. 
    However, both features may be influenced by systematic effects, and further investigation is required.}
    \label{fig:A4}
\end{figure*}

We further examine the age distribution in the [Mg/Fe]--[Fe/H] plane by separating stars into mono-abundance populations. 
\autoref{fig:A2} shows histograms of ensemble kinematic ages in bins of [Fe/H] and [Mg/Fe] with widths of 0.1~dex. 
For each bin, we perform Gaussian mixture modeling (GMM), with the number of components determined using the Bayesian information criterion (BIC). 
The black curves show the composite distributions, while the colored histograms indicate the decomposed components. 
The peak locations and their relative weights for the older and younger populations are summarized in \autoref{fig:A4} at the corresponding [Mg/Fe]--[Fe/H] bin centers. 
We display only peaks separated by more than 3~Gyr.

Interestingly, most bins exhibit bimodal age distributions. 
Combined with the age--[Mg/Fe] and age--[Fe/H] relations shown in the top row of \autoref{fig:A4}, we find that the bimodality in the [Mg/Fe]--[Fe/H] plane corresponds to the two populations seen in the age--[Mg/Fe] relation at ages $> 8$~Gyr. 
The older peak of the bimodal age distribution in the low-$\alpha$ disk corresponds to the younger tail of the population in the age--[Mg/Fe] plane at ages $> 12$~Gyr, where [Mg/Fe] decreases from $\sim 0.2$~dex to $< 0$~dex. 

The younger peak of the bimodal age distribution in the high-$\alpha$ disk corresponds to the younger tail of the population in the age--[Mg/Fe] plane at ages between 7 and 12~Gyr, where [Mg/Fe] decreases to $\sim 0.2$~dex and then increases again to $\sim 0.3$~dex at $\sim 10$~Gyr. 
Further inspection shows that this upturn occurs primarily among inner-disk giant stars with guiding radii $< 6$~kpc. 
However, both features may be influenced by systematic effects, including biases in ensemble kinematic ages and abundance measurements; therefore, additional investigation is required. 
Excluding stars with [Fe/H] $< -0.5$~dex, as recommended based on tests with subgiant ages (see \autoref{sec:kinage} and \autoref{fig:fig3}), does not qualitatively alter these results.

\bibliography{sample631}{}
\bibliographystyle{aasjournal}



\end{document}